\documentclass[]{aa}

\usepackage{graphicx}
\usepackage{amssymb}
\usepackage{amsmath}
\usepackage{multirow,bigdelim}
\usepackage{txfonts}
\usepackage[export]{adjustbox}
\usepackage{xcolor}

\usepackage{hyperref}
\hypersetup{
    colorlinks=true,
    linkcolor=blue,
    citecolor=blue,
    urlcolor=blue
}

\begin{document} 

   \title{Radio view of the Peanut cluster: merger-induced stripping  and a radio halo shortly after pericentric passage}

\authorrunning{Rajpurohit et al.}
\titlerunning{Radio view of the Peanut cluster}

\author{
K. Rajpurohit\inst{1}
\and
N. Lyskova\inst{2}
\and
E. Churazov\inst{2}
\and
W. Forman\inst{1}
\and
I. Khabibullin\inst{3, 2}
\and
R. Kraft\inst{1}
\and
L.~Di Mascolo\inst{4}
\and
T.~Mroczkowski\inst{5, 6}
\and
A.~Botteon\inst{7}
\and
C. Jones\inst{1}
\and
E. O'Sullivan\inst{1}
\and
C. Zhang\inst{9, 10}
\and
R. J. van Weeren \inst{8}
\and
I. Zhuravleva\inst{10}
\and
G. Schellenberger\inst{1} 
\and
J. M. Vrtilek\inst{1}
\and
L. P. David\inst{1}
\and
R.~Sunyaev\inst{2}
\and 
M.~Devlin\inst{11}
\and
S.~Dicker\inst{11}
\and 
S.~Haridas\inst{11}
\and
J.~van~Marrewijk\inst{8}
\and
B.~S.~Mason\inst{13}
\and
E.~Moravec\inst{14}
\and
J.~Orlowski-Scherer\inst{11}
\and
C.~Romero\inst{11, 12}
\and
C.~L.~Sarazin\inst{15}
\and
J.~Sievers\inst{16} 
}

\institute{
Center for Astrophysics | Harvard \& Smithsonian,
60 Garden St., Cambridge, MA 02138, USA
\and
Max Planck Institute for Astrophysics, Karl-Schwarzschild-Str. 1, Garching, D-85741, Germany
\and
Rudolf Peierls Centre for Theoretical Physics, Department of Physics, University of Oxford, Clarendon Laboratory, Parks Rd,
Oxford, OX1 3PU, United Kingdom
\and
Kapteyn Astronomical Institute, University of Groningen, Landleven 12, NL-9747 AD, Groningen, The Netherlands
\and
Institute of Space Sciences (ICE-CSIC), Carrer de Can Magrans, s/n, 08193 Cerdanyola del Vall\`es, Barcelona, Spain
\and
Institut d'Estudis Espacials de Catalunya (IEEC), E-08860 Castelldefels, Barcelona, Spain
\and
INAF - IRA, via P. Gobetti 101, 40129 Bologna, Italy
\and
Leiden Observatory, Leiden University, PO Box 9513, 2300 RA Leiden, The Netherlands
\and
Department of Theoretical Physics and Astrophysics, Faculty of Science, Masaryk University, Kotl\'{a}\v{r}sk\'{a} 2, Brno 611 37, Czech Republic
\and
University of Chicago, 5801 South Ellis Avenue, Chicago, IL, 60637, USA
\and
University of Pennsylvania, 209 S. 33rd St., Philadelphia, PA 19104, USA
\and
ADNET Systems, Inc., 6720B Rockledge Drive, Bethesda, MD 20817, USA
\and
National Radio Astronomy Observatory, 520 Edgemont Rd., Charlottesville, VA 22903, USA
\and 
Green Bank Observatory, P.O. Box 2, Green Bank, WV 24944, USA
\and
Department of Astronomy, University of Virginia, P.O. Box 400325, Charlottesville, VA 22904, USA
\and
Department of Physics, McGill University, 3600 University Street, Montreal, QC, H3A 2T8, Canada
}

  \abstract
{The Peanut galaxy cluster (SRGe CL0238.3+2005) offers a rare opportunity to investigate thermal and nonthermal processes only $\sim0.1$~Gyr after pericentric passage. We present a multi-frequency radio study of this system at $z=0.42$ using new MeerKAT and uGMRT observations, complemented by \textit{Chandra} X-ray and MUSTANG-2 data. We identify two morphologically and spectrally distinct radio components: a double-lobed source associated with the northern subcluster and extended diffuse emission distributed across the central cluster region. The double-lobed source has a strongly curved, ultra-steep spectrum and is offset by $\sim22$~kpc from the nearest brightest cluster galaxy. Its spectral curvature and positional offset are consistent with remnant AGN plasma that may have been displaced during the merger or reshaped by a shock. The diffuse emission extends over $\sim1$~Mpc and is classified as an ultra-steep spectrum ($\alpha_{\rm 400~MHz}^{\rm 1.28~GHz}=-1.77\pm0.11$) radio halo. It has a volume-averaged synchrotron emissivity of $(1.1\pm0.1)\times10^{-41}\,{\rm erg\,s^{-1}\,cm^{-3}\,Hz^{-1}}$ at 1.4~GHz,  an order of magnitude higher than the characteristic value reported for other radio halos and twice that of the Bullet cluster halo. The radial profile of the halo is well described by a single exponential fit and its radio emission exhibits a tight spatial correlation with the X-ray surface brightness, with a best-fit slope of $1.11\pm0.05$. \textit{Chandra} and MUSTANG-2 observations reveal candidate surface brightness discontinuities, with the most prominent inner feature coinciding with a region of flatter radio spectral index, potentially tracing localized particle (re)-acceleration. The coexistence of an extended ultra-steep spectrum radio halo, high synchrotron emissivity, remnant AGN plasma, and candidate X-ray/SZ discontinuities makes the Peanut cluster an important test case for understanding the evolution of diffuse radio emission and its connection to the dynamical state of the intracluster medium during the early stages of cluster mergers.}

   \maketitle

\section{Introduction}
\label{introduction}

Galaxy clusters grow through hierarchical accretion and mergers, which are among the most energetic events in the Universe. These mergers profoundly affect the intracluster medium (ICM), driving shocks, turbulence, and large scale gas motions, ultimately accelerating relativistic particles. Systems observed shortly after pericentric passage are rare because this phase is short-lived. The Peanut cluster (SRGe~CL0238.3+2005) was recently identified in the all-sky survey conducted with the extended ROentgen Survey with an Imaging Telescope Array (eROSITA) aboard the Spectrum-Roentgen-Gamma (SRG) observatory \citep{2021A&A...656A.132S, 2021A&A...647A...1P, 2022AstL...48..702B, Lyskova2025}. Located at $z=0.42$, the cluster has a mass of $M_{500}\simeq10^{15}\,M_\odot$ and an ICM temperature of $kT\simeq10$~keV. Multi-wavelength observations indicate that the system is undergoing a major merger, approximately $0.1$~Gyr after pericentric passage \citep{Lyskova2025,2026JHEAp..5300632Z}. Such massive systems at this redshift are rare and only a few dozen galaxy clusters with $M_{500} > 10^{15}\,M_\odot$ are expected to be found across the whole sky \citep{2008ApJ...688..709T, 2011PhRvD..83b3015M, 2013JCAP...07..022H}. 

The \textit{Chandra} 40-ks X-ray image of the Peanut cluster reveals two bright X-ray clumps to the north and south, separated by a pronounced surface brightness depression \citep{Lyskova2025}. Comparison with optical observations shows that the X-ray bright gas component is offset from the subcluster galaxies, reminiscent of the Bullet cluster merger  \citep[see, e.g.,][]{2004ApJ...606..819M,2006A&A...451..395C,2016A&A...594A.121P}. In contrast to the Bullet cluster, however, the Peanut cluster appears to involve two ``bullets'' in an almost head-on collision. \cite{Lyskova2025} reported that the cluster experienced an almost head-on merger only $\lesssim 0.1\,{\rm Gyr}$ ago, i.e., it is observed in a particularly energetic phase of the merger (just after the pericenter passage of the two merging sub-halos).

\begin{table*}[!thbp] 
\centering
\caption{Observational overview: MeerKAT, uGMRT and LOFAR}
\label{tab::obs}
\begin{tabular}{lccccc}
\hline\hline
& MeerKAT L-band & MeerKAT UHF-band & uGMRT Band~3 & uGMRT Band~4  & LOFAR HBA \\
\hline
Frequency range      & 0.9-1.7~GHz & 0.55-0.9~GHz & 300-500~MHz & 550-850~MHz &120-169~MHz \\
Channel width        & 208~kHz & 132.81~kHz &  97.7~kHz& 42~kHz&390.6 kHz \\
No.\ of channels     &  4028& 4028 &  2048& 4096 &125\\
On-source time       & 3 hr& 3 hr &6 hr & 6 hr & 8 hr\\
\hline
\end{tabular}
\end{table*}

In the hierarchical build-up model \citep[e.g.,][]{2005Natur.435..629S,2005RvMP...77..207V, 2012ARA&A..50..353K}, major mergers are expected to play a key role in the growth of the most massive galaxy clusters. Systems observed shortly after core passage are of particular interest because they provide a unique opportunity to study a range of physics, from the dynamics of mergers and the nature of (self-interacting) dark matter \citep[e.g.,][for a recent review]{2025RvMP...97d5004A} to plasma effects in the ICM associated with shock waves, including electron-ion equilibration efficiency \citep{2019A&A...628A.100D, 2022MNRAS.514.1477R} and particle acceleration \citep{Botteon2026}. 

Diffuse radio emission associated with galaxy clusters, known as radio halos and radio relics, provides a direct probe of the nonthermal processes in the ICM \citep{2007PhR...443....1M,vanWeeren2019, 2025A&A...694A.182A}. These sources and fossil plasma from active galactic nuclei (AGN) \citep[e.g., ][]{Mandal2024,Brienza2022, Botteon2026b, Rajpurohit2024a} trace the distribution of relativistic particles and magnetic fields, (re)-accelerated by processes like shocks and turbulence. The Peanut cluster offers a rare opportunity to investigate diffuse radio emission in a massive cluster merger shortly after pericentric passage. However, despite its unusual dynamical state, the system remains poorly studied at radio wavelengths. Survey observations, including the Rapid Australian Square Kilometre Array Pathfinder (ASKAP) Continuum Survey \citep[RACS;][]{McConnell2020} and the Giant Metrewave Radio Telescope (GMRT) Sky Survey \citep[TGSS;][]{Intema2017} presented by \citet{Lyskova2025} revealed extended steep spectrum radio emission, potentially indicative of aged radio plasma. However, the sensitivity and angular resolution were insufficient to establish the nature of radio emission or to investigate its spectral properties in detail.

In this paper, we present new multi-frequency radio observations and a spectral analysis of the Peanut cluster. We use new radio observations conducted with MeerKAT \citep{Jonas2916}, the upgraded GMRT (uGMRT), and the archival LOw-Frequency ARray \citep[LOFAR;][]{Haarlem2013} High-Band Antenna. The new observations were proposed to characterize the radio emission in this system, enabling us to investigate its origin. These observations are complemented by new MUSTANG-2 observations of the thermal Sunyaev-Zeldovich signal and archival \textit{Chandra} X-ray data, which together probe the thermal and nonthermal components of the cluster.

The paper is structured as follows: Section~\ref{sec::observations} provides an overview of the observations and details of the data reduction. Section~\ref{sec::results} presents the newly obtained continuum images. The analysis is presented in Section~\ref{sec::analysis} followed by a discussion in Section~\ref{sec::discussion}.
A summary of the findings is provided in Section~\ref{sec::summary}.

Throughout this paper, we adopt a flat $\Lambda$CDM cosmology with $H_{\rm{ 0}}=70$ km s$^{-1}$\,Mpc$^{-1}$, $\Omega_{\rm{ m}}=0.3$, and $\Omega_{\Lambda}=0.7$. At the cluster's redshift, $z=0.42$, $1\arcsec$ corresponds to a physical scale of 5.54\,kpc. All radio images are in the J2000 coordinate system and corrected for primary beam attenuation. The radio spectral index is defined as $S_{\nu}\propto \nu^{\alpha}$, where $S_{\nu}$ is the flux density at frequency $\nu$.

\begin{figure*}[!thbp] 
    \centering
    \includegraphics[width=0.66\textwidth]{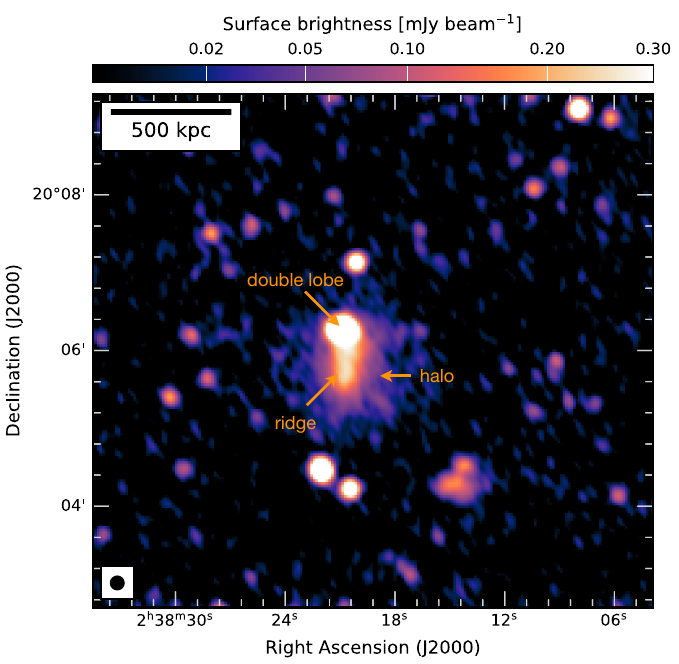}
    \vspace{-0.3cm}
 \caption{MeerKAT L-band (0.9-1.7\,GHz) image of the Peanut cluster with an angular resolution of 10\arcsec, displayed using a square-root intensity scale. The main features are labeled. The image reveals extended diffuse radio emission at the cluster center and a bright source to the north. The image was produced using Briggs weighting (robust=$-0.2$) and uv-tapering. The rms noise level is $11~\mu{\rm Jy~beam^{-1}}$. The beam size is indicated in the bottom left corner of the image.}
      \label{fig::labelling}
\end{figure*}

\section{Observations and data reduction}
\label{sec::observations}
We observed the Peanut cluster with the MeerKAT, uGMRT, and  Green Bank Telescope (GBT). These new radio observations were combined with the archival \textit{Chandra} X-ray data presented in \cite{Lyskova2025}. We also use the LOFAR Two-metre Sky Survey (LoTSS) data \citep{Shimwell2022, Shimwell2019, Shimwell2026}. An overview of the observations is given in Table\,\ref{tab::obs}.

\subsection{MeerKAT}

The cluster was observed with MeerKAT at L-band (0.9-1.7~GHz) and UHF band (0.55-0.9~GHz). The observations were carried out on March 8, 2026 (Proposal ID: SCI-20260306).  For both bands, J0408-6545 and J0238+1636 were used as the flux/bandpass and phase calibrators, respectively. J0521+1638 was included as a polarization calibrator with multiple scans. All four polarization products (XX, XY, YX, and YY) were recorded using the 4K correlator mode. The total recorded bandwidth was 856~MHz and 544~MHz at L-band and  UHF band, respectively.

The data were calibrated using the Containerized Automated Radio Astronomy Calibration \citep[{\tt CARACal};][]{caracal2020}. The initial processing in {\tt CARACal} included flagging of shadowed antennas, auto-correlations, and channels contaminated by known radio frequency interference (RFI) using the {\tt tfcrop} algorithm. The frequency range covering Galactic emission was flagged. Subsequently, flagging using {\tt AOFlagger} \citep{Offringa2010} was done to identify and remove low-level RFI. {\tt CARACal} modeled the primary calibrator J0408-6545 using the MeerKAT local sky models. This was followed by cross-calibration to determine the time-dependent antenna delays, complex gain solutions, and bandpass corrections.  Following the initial calibration, an image of the target field was generated using {\tt WSClean} \citep{Offringa2014}. This was followed by three rounds of phase-only self-calibration using {\tt CubiCal} \citep{Kenyon2018}, and a final round of amplitude-phase self-calibration. Polarization calibration was performed using the {\tt CARACal} {\tt polcal} worker, with an unpolarized calibrator for leakage calibration and J0521+1638 as the polarized calibrator to solve for the cross-hand phase and delay terms. In the end, all the solutions were applied to the target. The calibrated measurement sets were binned in frequency by a factor of 10 and imaged in {\tt WSClean} using the Briggs weighting scheme with ${\tt robust}=0$ and multiscale cleaning.

\subsection{uGMRT}

We also observed the Peanut cluster with the uGMRT in Band~3
and Band~4 (Project code: 48\_123) using the GMRT Wideband Backend. 3C~48 and 3C~147 were included as flux and bandpass calibrators. The initial calibration of the wideband uGMRT data was carried out using the Source Peeling and Atmospheric Modeling \citep[{\tt SPAM};][]{Intema2009} pipeline. For details of the primary data reduction, we refer to \cite{Rajpurohit2021c}. In summary, the Band~3 and Band~4 datasets were divided into six subbands each. The flux density scale was set using 3C48 following \cite{Scaife2012}. The data were then averaged, flagged, and corrected for the bandpass response. Subsequently, the calibrated visibilities were visually inspected for residual RFI contamination, and the affected data were further flagged.

To refine the phase calibration of the target field, we used an initial global sky model derived from the GMRT Software Backend (GSB). The SPAM-calibrated subbands were then imaged in WSClean using Briggs weighting with a robust parameter of 0 to create full band continuum images. The deconvolution was always performed using multiscale cleaning \citep{Offringa2014}. 

\subsection{LOFAR}

The cluster was observed with LOFAR HBA as part of the
LOFAR Two-metre Sky Survey (LoTSS). The observations were conducted in HBA dual inner mode. The data reduction and calibration were performed using the standard pipelines used by the LOFAR Surveys Key Science Project team \citep{2021A&A...648A...1T,Williams2016, vanWeeren2016a,deGasperin2019,Shimwell2026}. The data were then extracted from the pointings P038+21, P040+19, and P041+21 and re-calibrated using facetselfcal \citep{2021A&A...651A.115V} to refine the calibration in a $45^\prime \times 45^\prime$ field containing the target and to allow for deeper deconvolution and faster re-imaging. For more details we refer the reader to \cite{2021A&A...651A.115V,2022A&A...660A..78B,2026MNRAS.546ag054V}.

\begin{figure*}[!thbp] 
    \centering
    \includegraphics[width=0.50\textwidth]{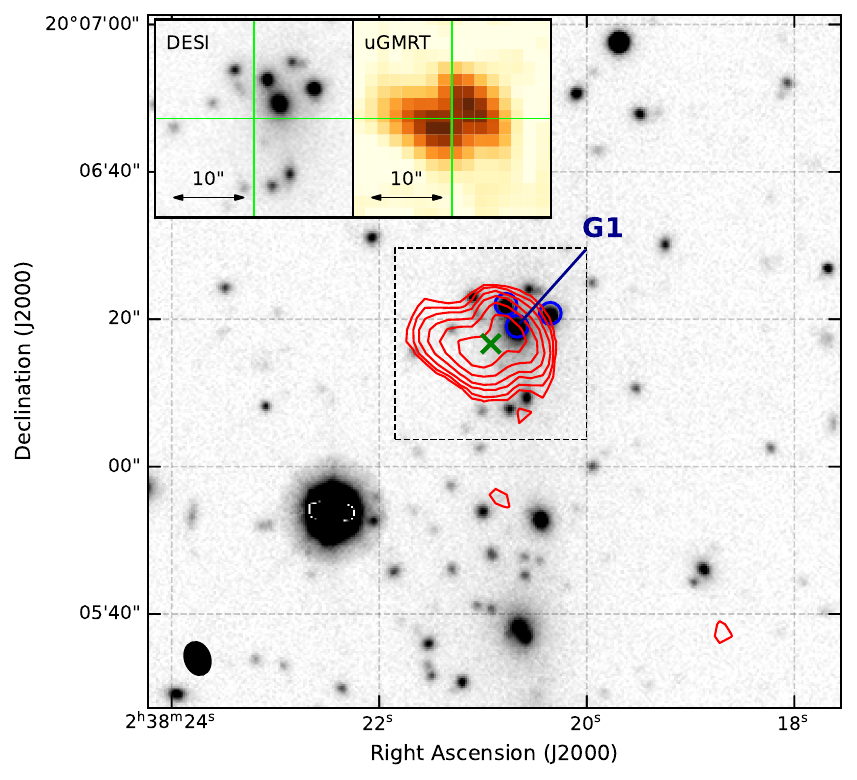}
        \includegraphics[width=0.485\textwidth]{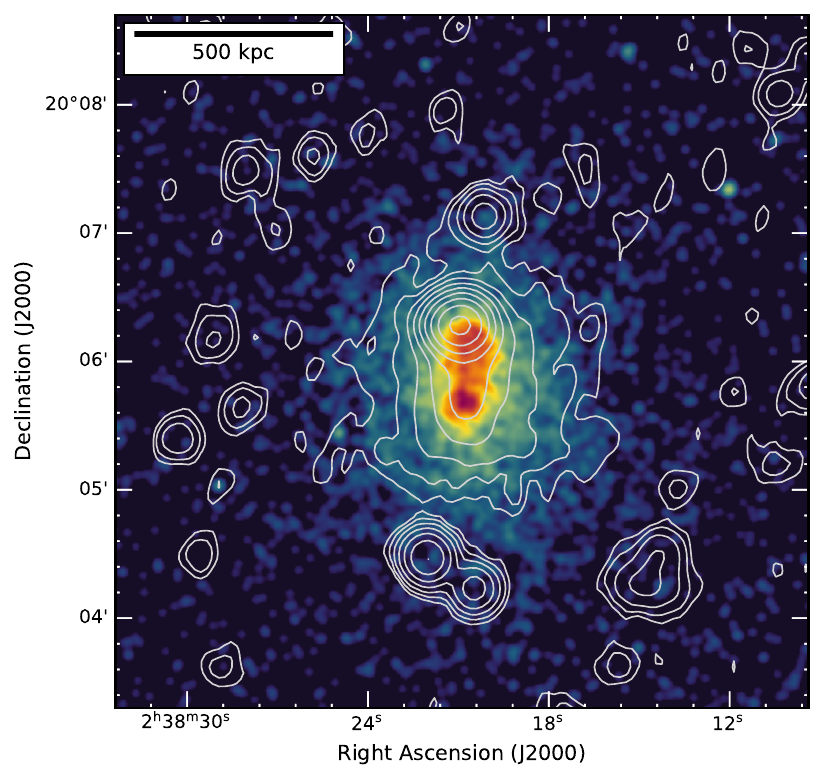}
 \caption{\textit{Left}: Dark Energy Spectroscopic Instrument (DESI) Legacy Survey DR10 $r+z$ image overlaid with the uGMRT Band4 (beam size: $4.6\arcsec \times 3.4\arcsec$) radio contours. The image highlights a double-lobed radio source to the north whose centroid (green cross mark) is not centered on the northern BCG (G1) or any other optical galaxy (marked with blue circles). The insets show a $26\arcsec\times26\arcsec$ zoom of the double-lobe source (dashed box) in the DESI (left) and uGMRT Band~4 (right) images. The uGMRT Band4 contour levels are drawn at  $[1, 2, 4, 8 ...]\times 4.0\sigma_{\rm rms}$. \textit{Right}: \textit{Chandra} (0.9-6 keV) X-ray image Gaussian smoothed with a FWHM of 3\arcsec\ and overlaid with a MeerKAT 815~MHz radio contours (15\arcsec\ resolution) drawn at  $[1, 2, 4, 8 ...]\times 3.0\sigma_{\rm rms}$ where rms is $16\,\mu{\rm Jy\,beam^{-1}}$. The image shows that the central diffuse emission broadly follows the X-ray emission. }
      \label{fig::optical_X-ray}
\end{figure*} 

\subsection{GBT/MUSTANG-2}

MUSTANG-2 is a 90 GHz broadband continuum bolometer camera \citep{Dicker2014, Stanchfield2016} installed on the Green Bank Telescope (GBT), offering an angular resolution of approximately $9''$ and a $4.2'$ diameter field of view. For massive clusters such as the Peanut, MUSTANG-2 primarily provides resolved measurements of the thermal Sunyaev-Zeldovich (SZ) effect \citep{1972CoASP...4..173S}, which traces the line of sight integrated electron pressure \citep[see also][for reviews]{1999PhR...310...97B, 2019SSRv..215...17M}.
MUSTANG-2 spent 30 h  observing the Peanut cluster (Project GBT25B-215, PI: Luca Di Mascolo). Our data reduction procedure followed the reduction and map making methodology detailed in \cite{Romero2020}. Detailed analysis of the MUSTANG-2 data will be presented in a forthcoming paper.

\section{Results: Morphology}
\label{sec::results}

Our MeerKAT L-band (0.9-1.7~GHz) radio continuum image of the Peanut cluster at 10\arcsec\ resolution is shown in Figure\,\ref{fig::labelling}. The new observations reveal previously undetected radio structures, providing new insight into the dynamical evolution of the system. The image reveals large scale diffuse radio emission. To facilitate the discussion, we label the most prominent features: the double-lobe source in the northern subcluster and the diffuse emission that fills the cluster center, the brightest part of which we refer to as the ridge. These features are described in the following subsections.

\begin{figure*}[!thbp] 
    \centering
    \includegraphics[width=0.90\textwidth]{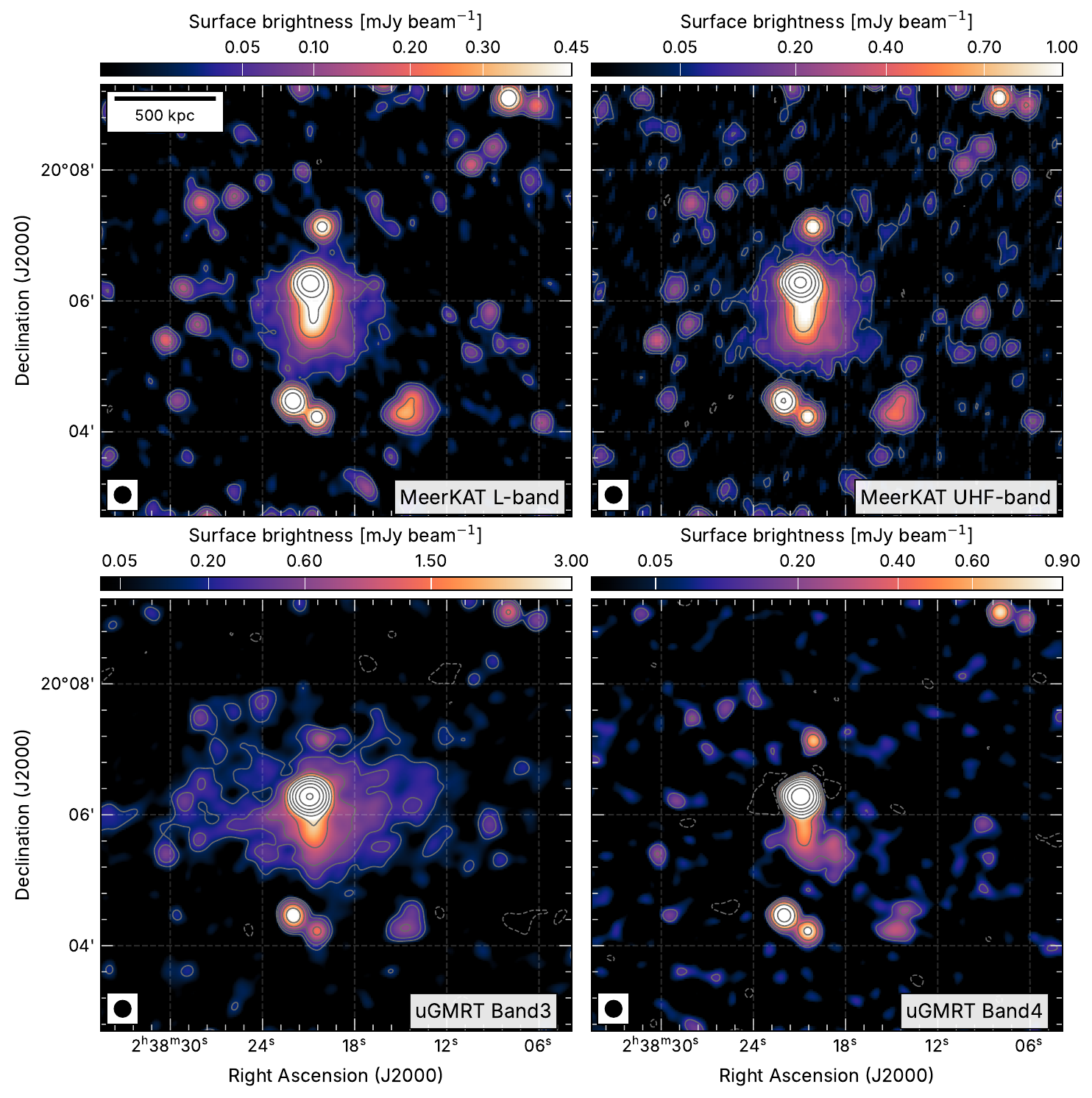}
    \vspace{-0.2cm}
 \caption{Peanut cluster radio images at a common resolution of 15\arcsec\ obtained from MeerKAT L-band (0.9-1.7~GHz), MeerKAT UHF-band (0.55-0.9 GHz), uGMRT Band~3 (300-500~MHz) and uGMRT Band~4 (550-850~MHz) in square-root scale. The images highlight the central diffuse radio emission. Radio contour levels are drawn at  $[1, 2, 4, 8 ...]\times 3.0\sigma_{\rm rms}$. Dashed contours depict the $-3.0\sigma_{\rm rms}$. All images were produced using Briggs weighting with a robust=0 and $uv$-tapering. The corresponding rms noise levels are $\sigma_{\rm L-band} = 10\,\mu{\rm Jy\,beam^{-1}}$, $\sigma_{\rm UHF-band} = 16\mu{\rm Jy\,beam^{-1}}$, $\sigma_{\rm Band\,3} = 50\,\mu{\rm Jy\,beam^{-1}}$, and $\sigma_{\rm Band\,4} = 34\,\mu{\rm Jy\,beam^{-1}}$.}
      \label{fig::15arcsecmaps}
\end{figure*}

\subsection{Double-lobe source}
The brightest source in the cluster region lies in the northern subcluster and is detected from 144~MHz to 1.28~GHz (labeled `double lobe'). A ridge of radio emission connects this source to the southern subcluster. At the $10\arcsec$ resolution of this image, the compact source and the adjoining ridge together resemble the head and the tail of a head-tail radio galaxy. However, the higher resolution images reveal a different morphology.

\begin{figure*}[!thbp] 
    \centering
    \includegraphics[height=7.8cm]{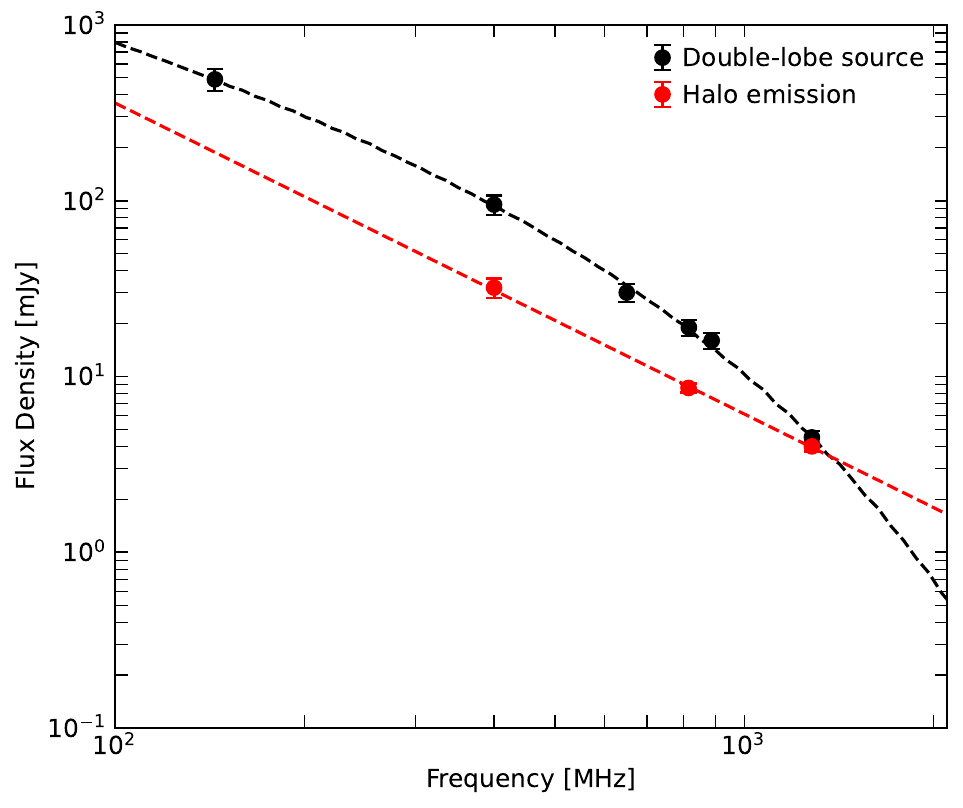}
    \hspace{0.5cm}
    \includegraphics[height=8.0cm]{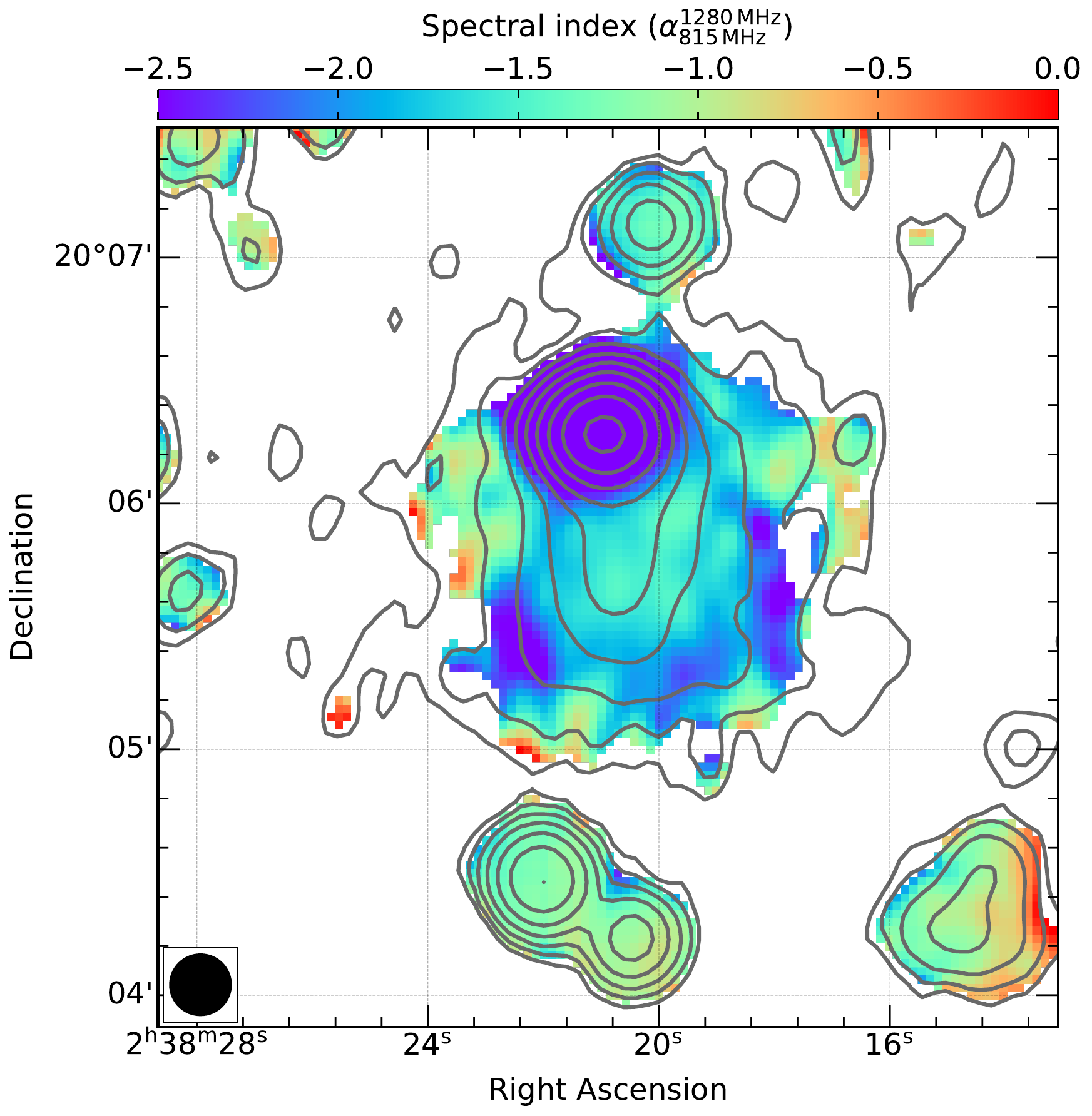}
    \caption{\textit{Left:} Integrated spectrum of the double-lobe source obtained from 144~MHz to 1.28~GHz and the central halo emission between 815~MHz and 1.28~GHz. The double lobe source exhibits a high frequency steepening and the spectrum is fitted  with a JP model (black dashed line) with an injection index of $-0.85$. The diffuse emission has a spectral slope of $-1.77\pm0.11$ with the red dashed line representing the power-law fit. \textit{Right:} Spectral index map created between 815~MHz and 1.28~GHz. The angular resolution is 15\arcsec. }
    \label{fig::spec}
\end{figure*}

\begin{table*}
\caption{Integrated flux densities of the double-lobe source and radio halo.}
\label{tab:fluxes}
\centering
\begin{tabular}{lcccccc}
\hline\hline
Source & $S_{144}$ & $S_{400}$ & $S_{650}$ & $S_{816}$ & $S_{887} (RACS)$ & $S_{1280}$ \\
       & (mJy)    & (mJy)     & (mJy)     & (mJy)     & (mJy)     & (mJy)      \\
\hline
Double-lobe  &
$490 \pm 70$ &
$95 \pm 12$ &
$30 \pm 3.5$ &
$19 \pm 2.0$ &
$16 \pm 1.7$ &
$4.5 \pm 0.4$ \\

Halo &
-- &
$32 \pm 4$ &
-- &
$8.6 \pm 0.5$ &
-- &
$4.0 \pm 0.24$ \\
\hline
\end{tabular}
\tablefoot{Flux density errors include flux density scale uncertainties which are assumed to be 10\% for LOFAR, uGMRT Band~3 and Band~4, and 5\% for MeerKAT L-band and UHF data. The flux densities of the double-lobe source and halo were measured from 15\arcsec\ resolution images, created using a common uv-cut. }
\end{table*}

In the left panel of Figure\,\ref{fig::optical_X-ray} we present the DESI $r+z$ image overlaid with the uGMRT Band~4 (550-850~MHz) contours, imaged at the highest resolution available for these data, $4.6\arcsec \times 3.4\arcsec$. The source is resolved into two lobes with a total angular extent of $\approx35\arcsec$, corresponding to $\approx190$~kpc in projection, favoring a double-lobed morphology over the head-tail interpretation suggested by the lower-resolution image. Circles mark three galaxies of the northern subcluster. All three are spectroscopically confirmed cluster members, at $z=0.4252\pm0.0004$, $0.4253\pm0.0005$, and $0.4258\pm0.0006$ \citep{Lyskova2025,2026JHEAp..5300632Z}. The brightest of them is G1. The radio centroid, marked by a cross, coincides with none of these galaxies and is offset from G1 by $\approx4\arcsec$, that is $\approx22$~kpc in projection. These galaxies are displaced from the X-ray peak of the northern subcluster \citep{Lyskova2025}. Relative to the X-ray peak, the centroid of the double-lobed radio source lies $\approx10\arcsec$ ($\approx50$~kpc in projection) to the northeast. The insets in the left panel of Figure\,\ref{fig::optical_X-ray} show zoomed-in views of the double-lobed source in the DESI and uGMRT Band~4 images, showing the offset more clearly. No compact radio core is detected within the source. 

\subsection{Diffuse radio emission}
Our new observations reveal large scale, low surface brightness diffuse radio emission for the first time, see Figures\,\ref{fig::labelling} and \ref{fig::15arcsecmaps}. At higher frequencies, the emission is elongated in the north-south direction, broadly aligned with the merger axis.  Its brightest part, the ridge, extends over $\approx400$~kpc southward from the northern subcluster, has a width of $\approx150$~kpc, and closely follows the chain-like distribution of optical galaxies. The diffuse emission extends beyond the double-lobe source in the northern subcluster. 

We present the \textit{Chandra} X-ray image (0.9-6~keV) of the cluster, smoothed with a Gaussian kernel of the full width at half maximum (FWHM) 3\arcsec, overlaid with the MeerKAT UHF contours at 15\arcsec\ resolution in Figure\,\ref{fig::optical_X-ray} right panel. The X-ray emission is resolved into two clumps separated by $\approx0.5\arcmin$, or $\approx165$~kpc in projection, with a depression between them where the surface brightness is $30-40\%$ lower than in the clumps \citep{Lyskova2025}. The diffuse radio emission follows the X-ray emission closely in both shape and extent.

In Figure\,\ref{fig::15arcsecmaps} we present multi-frequency radio maps of the cluster at a common resolution of 15\arcsec, obtained with MeerKAT at L-band and UHF and with the uGMRT at Band~3 and Band~4, covering the frequency range from 300~MHz to 1.7~GHz. The central diffuse emission is detected at all four frequencies. In contrast to the high frequency, the uGMRT Band~3 image reveals more extended emission with a complex morphology, including an apparent east-west extension. The diffuse emission has a largest linear size (LLS) of $\approx1$~Mpc in the uGMRT Band~3 image and $\approx800$~kpc in the MeerKAT UHF image, measured within the $3\sigma_{\rm rms}$ contours of the respective 15\arcsec\ resolution images.

The diffuse emission is best recovered in the MeerKAT UHF and uGMRT Band~3 images. Although uGMRT Band~4 and MeerKAT UHF cover similar frequency ranges, significantly less diffuse emission is recovered in Band~4, where only the brighter ridge is detected. This difference is primarily due to insufficient short-baseline coverage in the uGMRT Band~4 observations. The diffuse emission is not detected in the LOFAR HBA image owing to poor data quality (Figure\,\ref{fig::HBA}). We do not detect filamentary or relic-like emission in the Peanut cluster, either in its central region or in its periphery. The arc-like feature reported $\approx5\arcmin$ east of the cluster center by \citet{Lyskova2025} is resolved into a tailed radio galaxy in our higher-resolution images (Appendix\,\ref{sec::sourceC}).

\section{Analysis}
\label{sec::analysis}

To study the spectral properties of the double-lobe source and the central diffuse emission, and their connection with the thermal emission, we combined our MeerKAT, uGMRT and LOFAR observations with our MUSTANG-2 data and with the published \textit{Chandra} X-ray observation, for the data reduction see \cite{Lyskova2025}. 

\subsection{Radio spectral analysis}
\label{s:spec}

The radio observations reported here were performed with three interferometric arrays, each with a different $uv$ coverage. To measure the flux densities, we produced images using Briggs weighting with a robust parameter of $-0.5$ and applied a common inner $uv$ cut of $350\lambda$, corresponding to the shortest well-sampled baseline of the uGMRT observations. This uv-cut ensures that the radio emission is recovered over comparable angular scales at all frequencies. The full extent of the diffuse emission is recovered with MeerKAT and uGMRT Band~3 images. For the central diffuse emission, we therefore use the L-band, UHF, and Band~3 images to measure the integrated flux densities and spectral indices. The measurements were performed within the $3\sigma_{\rm rms}$ contour, excluding the emission associated with the double-lobe source. The flux densities of both components are listed in Table\,\ref{tab:fluxes}.

\begin{figure*}[!thbp] 
    \centering
    \includegraphics[width=0.92\textwidth]{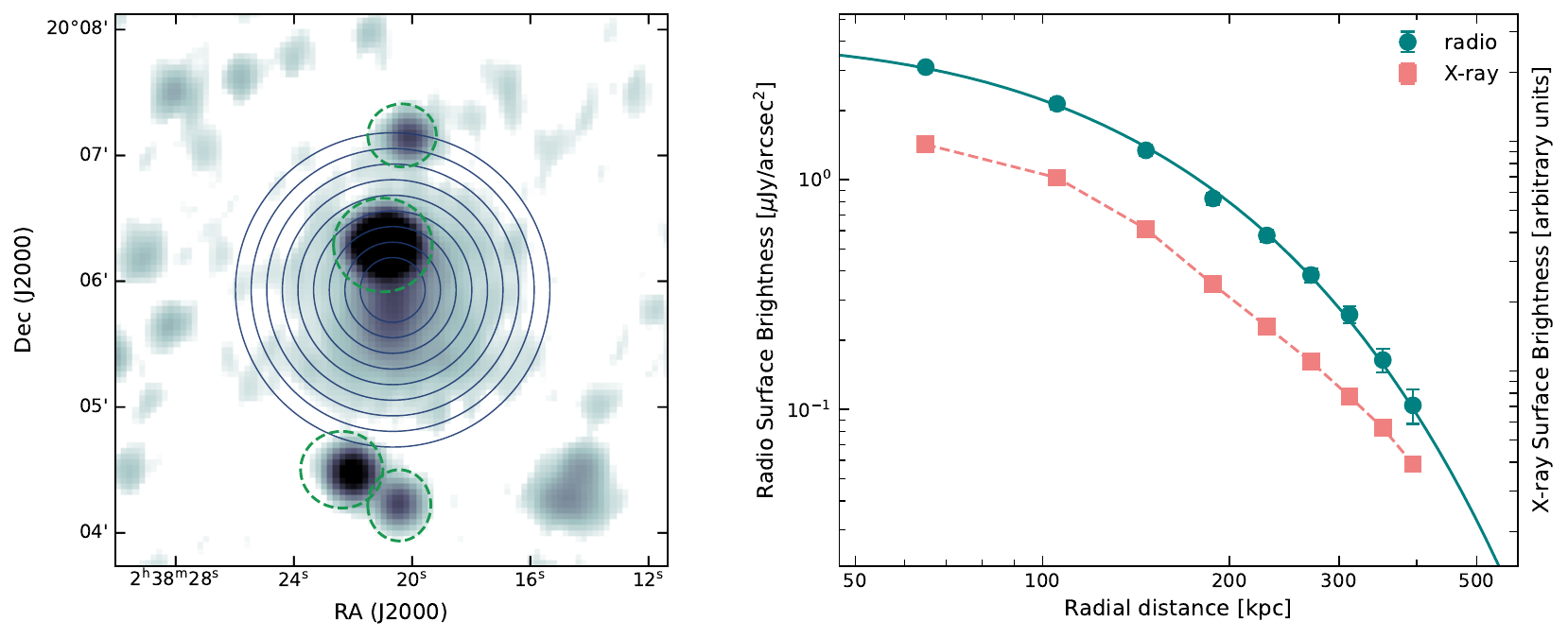}
 \caption{{\textit{Left:} MeerKAT UHF image at 15\arcsec\ resolution overlaid with the annuli used to extract the radial profile. The dashed circles mark the sources masked before the profile extraction, namely the double-lobe source and three compact sources. \textit{Right:} azimuthally averaged radial profile of the central diffuse emission at 815~MHz  and of the \textit{Chandra} surface brightness extracted in the same annuli. The solid line is the exponential fit to the radio radial profile. The width of the annuli is half of the beam FWHM, i.e., 7.5\arcsec. The central diffuse emission is well described by a single component.}}
    \label{fig::profile}
\end{figure*}

The flux density scale of all radio observations (MeerKAT, uGMRT and LOFAR) was checked by comparing the spectra of point-like discrete sources from 144~MHz to 1.28~GHz. We estimated the uncertainty in the flux density as follows:
\begin{equation}
\Delta S =  \sqrt {(f \cdot S)^{2}+{N}_{{\rm{ beams}}}\ (\sigma_{{\rm{rms}}})^{2}},
\end{equation}
where $f$ is the absolute flux density calibration uncertainty, $S$ is the flux density, $\sigma_{{\rm{ rms}}}$ is the noise level, and $N_{{\rm{beams}}}$ is the number of beams. We assumed absolute flux density uncertainties of 10\% for LOFAR \citep{Shimwell2026} and uGMRT \citep{Chandra2017} data and  5\% for MeerKAT UHF and L-band data.

In the left panel of Figure\,\ref{fig::spec} we show the integrated spectrum of the double-lobe source from 144~MHz to 1.28~GHz. It combines the LOFAR HBA, uGMRT Band~3 and Band~4, and MeerKAT UHF and L-band measurements with the RACS flux density at 887~MHz \citep{McConnell2020}. The source exhibits significant high-frequency spectral steepening, from $\rm \alpha_{144~MHz}^{650~MHz}=-1.85\pm0.03$ to $\rm \alpha_{815~MHz}^{1.28~GHz}=-3.20\pm0.02$. We fit the spectrum with a Jaffe-Perola \citep[JP;][]{JaffePerola1973} synchrotron aging model using the {\tt synchrofit} package. The best-fit injection index is $-0.85$, with a break frequency $\nu_{\rm break}=581^{+244}_{-172}~\rm MHz$. 

At 400~MHz, the newly detected diffuse emission accounts for approximately 25\% of the total flux density in the cluster region. Its integrated spectrum is well described by a power law with a spectral index of $-1.77\pm0.11$ between 400~MHz and 1.28~GHz. The diffuse emission is not detected in the LOFAR HBA image, while only the brighter ridge is recovered in uGMRT Band~4. These data are therefore excluded from the spectral analysis of the halo. Unlike the double-lobed source, the diffuse emission shows no evidence of spectral steepening over the frequency range reported. 

In the right panel of Figure\,\ref{fig::spec}, we present the spatially resolved spectral index map at $15\arcsec$ resolution between 815~MHz and 1.28~GHz. To construct the spectral index map, both L and UHF band images were re-gridded using the {\tt CASA} task {\tt imregrid}. The resulting images had slightly different resolutions, and we therefore convolved them to a common beam size of $15\arcsec\times15\arcsec$. Pixels below $3\sigma$ in either image were blanked. The corresponding error map is shown in Figure\,\ref{fig::spec_error}. The double-lobe source has the steepest spectral index in the cluster region with a mean value of $-3.0$ between 815~MHz and 1.28~GHz, whereas the diffuse emission exhibits a flatter spectral index of $-1.65$, consistent with the integrated spectral index values. 

We estimated the radio power of the diffuse emission using
\begin{equation}
P = 4\pi D_{\rm L}^{2} (1+z)^{-(1+\alpha)} S_\nu ,
\end{equation}
where $D_{\rm L}$ is the luminosity distance to the source, $\alpha$ is the spectral index, and $S_\nu$ is the measured flux density at 1.28~GHz. The factor $(1+z)^{-(1+\alpha)}$ accounts for the $k$-correction. For the measured flux density of $S_{\rm 1.28~GHz}=4.0\pm0.3$ mJy and $\alpha=-1.77\pm0.11$, we obtained a radio power of $P_{\rm 1.4~GHz} = (3.2 \pm 0.2)\times 10^{24}~{\rm W\,Hz^{-1}}$. Given the cluster mass of the order of $10^{15}M_{\odot}$, the radio power is consistent with the known $P_{\rm 1.4\,GHz}$ versus cluster mass scaling relation for radio halos \citep[e.g.,][]{Cuciti2021, Duchesne2024}.

The radial surface brightness profile of the central diffuse emission is shown in the right panel of Figure\,\ref{fig::profile}. The profile was extracted in concentric annuli centered on the X-ray centroid, with the width of each annulus set to half the full width at half maximum of the beam. The double-lobe source and other compact sources embedded in the halo were masked out from the map (Figure\,\ref{fig::profile} left panel). We included only those annuli with an average surface brightness above $3\sigma_{\rm rms}$. The profile is consistent with a single-component exponential of the form
\begin{equation}
I(r) = I_{0}\,e^{-r/r_{e}},
\end{equation}
where $I(r)$ is the azimuthally averaged surface brightness at a projected distance $r$ from the center, $I_{0}$ is the central surface brightness, and $r_{e}$ is the e-folding radius, that is the radius at which the brightness drops by a factor $e$. The best fit gives $I_{0}=6.5\pm0.3\,\mu{\rm Jy\,arcsec^{-2}}$ and $r_{e}=93\pm2$~kpc, as found for other radio halos \citep{Murgia2009, Bonafede2022, Rajpurohit2025}. The \textit{Chandra} surface brightness extracted in the same annuli declines at a similar rate over the same radial range, as found in other clusters hosting radio halos \citep[e.g., ][]{Govoni2001, Rajpurohit2021c}.

\begin{figure}[!thbp] 
    \centering
    \includegraphics[width=0.49\textwidth]{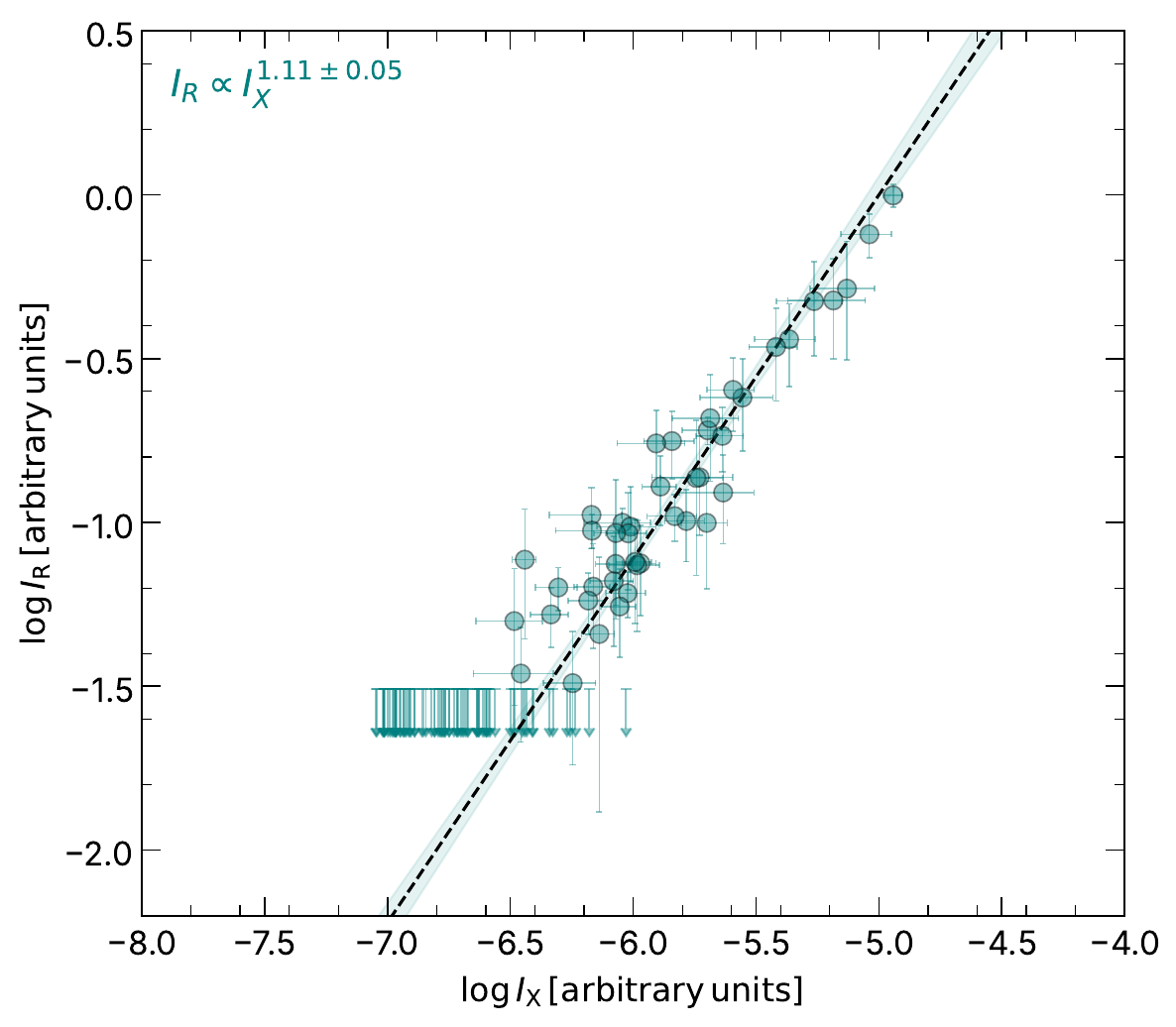}
    \caption{$\rm I_{R}-I_{X}$ relation of the radio halo in the Peanut cluster, extracted in square boxes with a width of $15\arcsec$ (corresponding to $\approx80$~kpc). The X-ray surface brightness was extracted from the \textit{Chandra} image smoothed with a Gaussian kernel of FWHM $15\arcsec$. The radio surface brightness was extracted from the MeerKAT UHF image at $15\arcsec$ resolution. The best-fit relation obtained using {\tt LinMix} is indicated by the dashed line. Arrows mark $2\sigma$ upper limits for cells with a radio surface brightness below $2\sigma$. The halo exhibits a near-linear correlation between the X-ray and radio surface brightness.}
    \label{fig::IRX}
\end{figure}

Following \citet{Murgia2009}, we estimated the average radio emissivity of the diffuse emission as
\begin{equation}
\langle J \rangle \simeq 7.7\times10^{-41}\,(1+z)^{3-\alpha}\,\frac{I_{0}}{r_{e}}
\ \ {\rm erg\,s^{-1}\,cm^{-3}\,Hz^{-1}},
\end{equation}
where $I_{0}$ is expressed in $\mu{\rm Jy\,arcsec^{-2}}$ and $r_{e}$ in kpc, and a filling factor of unity is assumed. For $I_{0}=6.5\pm0.3\,\mu{\rm Jy\,arcsec^{-2}}$ and $r_{e}=93\pm2$~kpc at 815~MHz, we obtain $\langle J\rangle = (2.8\pm0.2)\times10^{-41}\,{\rm erg\,s^{-1}\,cm^{-3}\,Hz^{-1}}$. Extrapolating $I_{0}$ to 1.4~GHz with the measured spectral index gives $I_{0}=2.6\pm0.1\,\mu{\rm Jy\,arcsec^{-2}}$ and $\langle J\rangle = (1.1\pm0.1)\times10^{-41}\,{\rm erg\,s^{-1}\,cm^{-3}\,Hz^{-1}}$. Radio halos exhibit a relatively narrow distribution of radio emissivity, with a characteristic value of $\sim10^{-42}\,{\rm erg\,s^{-1}\,cm^{-3}\,Hz^{-1}}$ \citep[e.g.,][]{Murgia2009, Rajpurohit2025}. In the $I_{0}-r_{e}$ plane used to compare radio halos \citep[see Figure~11 of][]{Rajpurohit2025}, the Peanut cluster lies close to the Bullet cluster, the most emissive halo of that sample, the two having comparable $I_{0}/r_{e}$. However, the volume-averaged radio emissivity of the Peanut cluster halo is approximately twice that of the Bullet cluster halo.

\begin{figure*}[!thbp] 
\begin{center}
\includegraphics[width=0.99\textwidth]{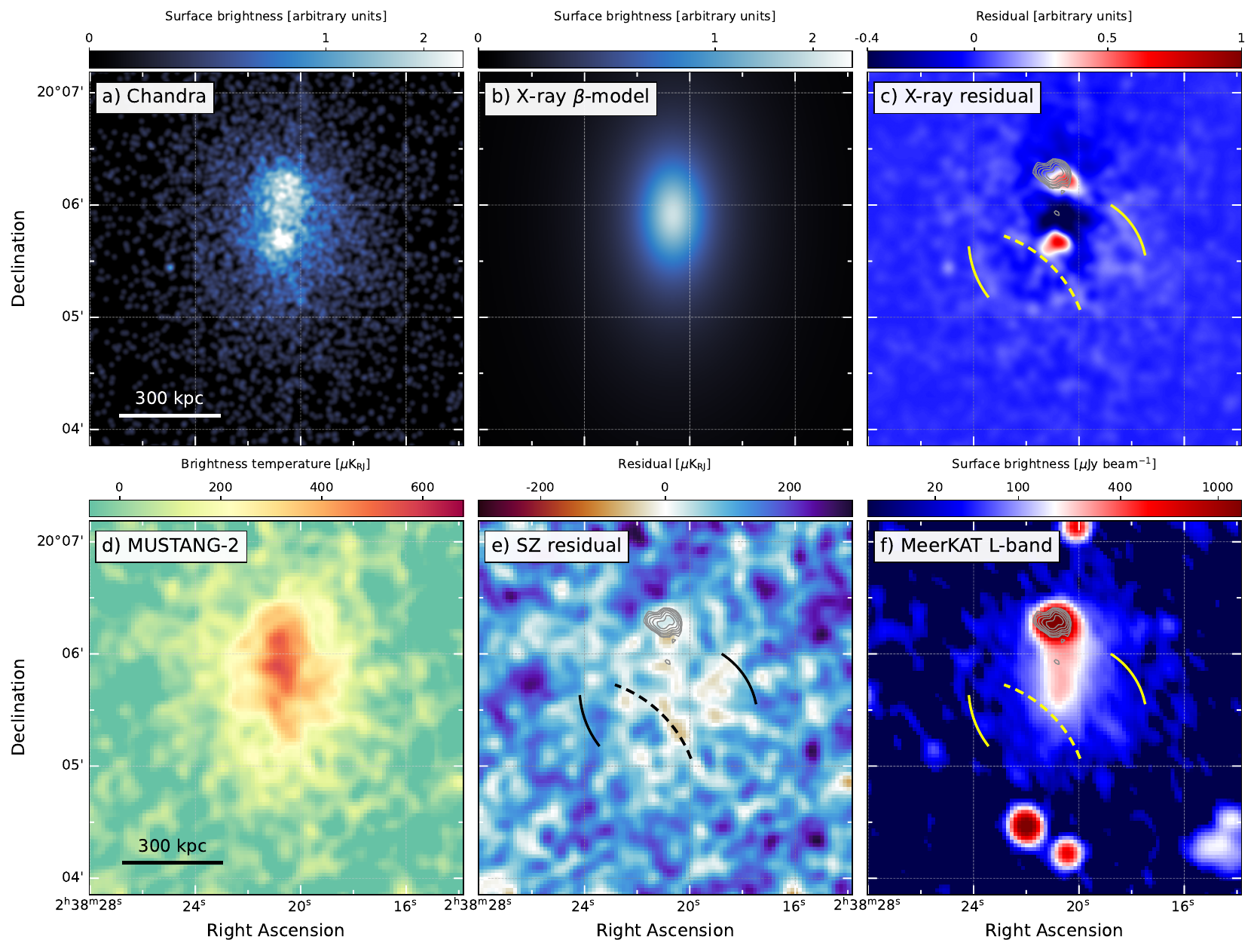}
\end{center}
\vspace{-0.5cm}
\caption{\textbf{Upper row, from left to right:} (a) \textit{Chandra} background-subtracted, exposure-corrected, 0.9-6.0~keV image of the Peanut cluster, smoothed by a 2 pixel Gaussian ($\approx1\arcsec$), (b) the elliptical $\beta$-model intended to approximate the overall surface brightness distribution of the cluster, and (c) the residual image obtained after subtraction of the model from the data, smoothed by a 5 pixel Gaussian ($\approx2.5\arcsec$). The image highlights deviations from the smooth gas distribution and candidate discontinuities (marked by lines). The position of the double-lobe radio source (gray contours) is also indicated. \textbf{Lower row, from left to right:} (d) the GBT/MUSTANG2 image of the Peanut cluster (in units of Rayleigh-Jeans brightness temperature $K_{\rm rj}$), smoothed by a 1.5 pixel Gaussian, resulting in an effective angular resolution of $\approx 11''$, (e) the residual image obtained after subtracting a best-fit elliptical $\beta$-model (similar to that in the upper row) from the data, and (f) the MeerKAT L-band image with an angular resolution of 10\arcsec. The most prominent candidate surface brightness discontinuity, the `inner' feature, appears as an arc-shaped enhancement (dashed line) in both the \textit{Chandra} and MUSTANG-2 residual images. It is also spatially coincident with a region of spectral index flattening (see Figure~\ref{fig::spec}).}  
\label{fig:shocks}
\end{figure*}

In Figure\,\ref{fig::IRX}, we present a point-to-point comparison between the radio and X-ray surface brightness across the central diffuse emission. The radio and X-ray surface brightness were extracted from the 15\arcsec\ resolution UHF image and the \textit{Chandra} image smoothed to a resolution of 15\arcsec, respectively. A grid of square cells, each 15\arcsec\ in size (i.e., similar to the radio beam size), was used to cover the diffuse emission region. The resulting grid is shown in Figure\,\ref{fig::IRX_cells}. Cells with both radio and X-ray emission above the $3\sigma$ threshold were included in the analysis. Cells with radio emission below $2\sigma$ were placed as upper limits. 

The radio and X-ray brightnesses are tightly correlated across the diffuse emission. Including upper limits, we obtain a best-fit slope of $1.11\pm0.05$, indicating a mildly superlinear relation. The fitting was performed in log-log space using the {\tt LinMix} package \citep{Kelly2007}, which accounts for measurement uncertainties on both axes, intrinsic scatter, and upper limits, following \cite{Botteon2020}. Restricting the fitting to $\geq3\sigma$ emission yields a slope of $0.93\pm0.06$, consistent with a linear relation.  For several radio halos, similar correlations have been reported, with either sub-linear or linear behavior \citep{Botteon2020, Rajpurohit2021c, Radiconi2022, Riseley2024, 2024A&A...686A...5B, Bonafede2022}.

\subsection{Candidate surface brightness discontinuities}
\label{sec::edges}

The \textit{Chandra} image of the Peanut cluster shows a complex and asymmetric X-ray morphology (Figure~\ref{fig:shocks} panel a). In addition to the elongation of the X-ray emission from north to south, localized structures are visible around the central region. The present data do not provide enough counts for spatially resolved spectroscopy, nor for fitting surface brightness profiles. We therefore searched for discontinuities in the residual images. 

We modeled the X-ray surface brightness distribution with a smooth elliptical $\beta$-model. As shown in Figure~\ref{fig:shocks} panel b, the model reproduces the overall elongation of the cluster emission, and was subtracted from the data. The residual image enhances departures from a smooth, relaxed gas distribution and reveals three coherent arc-like features around the cluster center, see Figure~\ref{fig:shocks} panel c. The most prominent, hereafter the `inner' arc, appears as a curved brightness excess southeast of the center (dashed line). Two fainter arcs, hereafter the `eastern' and `western' arcs, lie on either side of the center (solid lines). The inner arc lies $\approx140$~kpc from the cluster center at its closest point and extends to $\approx300$~kpc at its southern end. 

We also applied the same procedure to the MUSTANG-2 observations, shown in Figure~\ref{fig:shocks} panel d.  The resulting SZ residual map is shown in Figure~\ref{fig:shocks} panel e. Although the map is relatively noisy, an enhancement is visible near the position of the inner candidate discontinuity identified in the \textit{Chandra} residual image. The other candidate features are less clearly recovered in the SZ residual map. We therefore regard the correspondence at the inner feature as suggestive, but not conclusive.

The spatial relation between these candidate features and the radio emission is shown in Figure~\ref{fig:shocks} panel f. The inner candidate discontinuity lies within the diffuse radio emission, while the outer features are located near its boundaries. In addition, the MeerKAT spectral index map between 815~MHz and 1.28~GHz (Figure~\ref{fig::spec}) shows a region of spectral flattening near the position of the inner candidate feature. Among the structures identified in the X-ray residual image, the inner feature shows the clearest correspondence between the X-ray residual, SZ residual, and radio spectral index structure.

There is also marginal evidence for an additional surface brightness discontinuity at the northern boundary of the northern X-ray component, hereafter the `northern' feature, which is not marked in Figure~\ref{fig:shocks}. The \textit{Chandra} surface brightness decreases sharply across this boundary. The temperature measurements indicate that the northern component is hotter than the central region \citep{Lyskova2025}. Because the present data do not constrain the temperature across this boundary, its interpretation as a shock remains tentative.

\section{Discussion}
\label{sec::discussion}

The radio features described in Sects.~\ref{sec::results} and \ref{sec::analysis} differ in their morphologies and spectral properties. However, it is possible that all of them are shaped by the ongoing merger in the Peanut cluster. In the following subsections, we discuss plausible scenarios for the formation of the detected radio structures.

\subsection{Compact, steep spectrum source in the northern clump}
\label{s:compact}

Morphologically, the compact source to the north resembles a classic radio double-lobe source, associated with the activity of a supermassive black hole that powers bipolar jets \citep[e.g.][]{2020NewAR..8801539H}. We do not find any evidence for a spectral index gradient from the double-lobe source along the ridge region. The lack of continuous spectral aging along the ridge region implies that the double-lobe source is not a head-tail radio galaxy. The absence of such a gradient also indicates that the ridge is not a continuation of the lobes but part of the diffuse emission. Assuming that it is indeed a typical double-lobe radio galaxy, the measured size of each lobe is $\approx80\times40$~kpc. The size of known radio lobes in massive clusters varies by a large factor \citep{2019A&A...622A..12H}, sometimes reaching 100-200 kpc across, as, e.g., in the Ophiuchus cluster \citep{2020ApJ...891....1G}. The observed size therefore does not exclude the double-lobe source scenario.

The existing radio data do not have sufficient angular resolution to map and characterize the internal structure of this source in detail. However, even with this limited resolution, it is clear that the source exhibits double-lobe morphology and the centroid of radio emission does not coincide with any optical galaxy (see Figure\,\ref{fig::optical_X-ray} left panel). The galaxy G1 in the northern clump is likely one of the BCGs in this merging system and a plausible candidate for generating this radio source. The radio centroid is offset $\sim 4''$ ($22\, \rm$~kpc in projection) from this galaxy.  Although \textit{Chandra} data do not show any evidence for X-ray flux from the BCG core\footnote{No excess flux above the background in the 1-5~keV band is observed in a $2''$ circle centered at the G1 galaxy, with $1\,\sigma$ uncertainty in the X-ray luminosity of $\sim 10^{42}$ erg s$^{-1}$.}. 

Another notable property of this radio source is its ultra-steep spectral index, which steepens from $\rm \alpha_{144~MHz}^{650~MHz}=-1.85\pm0.03$ to $\rm \alpha_{815~MHz}^{1.28~GHz}=-3.20\pm0.02$.  The time required for the spectrum to steepen from $\alpha=-0.85$ (injection spectral index) to $\alpha\approx-1.8$  or $\alpha\approx -3.0$ due to radiative losses (synchrotron and inverse-Compton, IC) in a constant magnetic field $B=B_{\rm CMB}/\sqrt{3}\approx 3.8\,{\rm \mu G}$ (at $z=0.42$) is  $\sim 60\,{\rm Myr}$. This sets the longest possible timescale for the passively fading radio emission. The shortest timescale (in the same purely radiative cooling regime) can be estimated by assuming the energy density of the magnetic field is of the order of the gas pressure. Based on $n_e$ and $kT$ from the X-ray data, the gas pressure in the core is $P_{\rm gas}\sim 7\times 10^{-10}\,{\rm erg\,cm^{-3}}$. For $B^2/8\pi=0.5 P_{\rm gas}$, $B\sim 70\,{\rm \mu G}$. In this case, the characteristic timescale is $\sim 4\,{\rm Myr}$.

The combination of the centroid offset and the steepness of the radio spectrum suggests that the double-lobe source consists of aged radio lobes produced by an AGN in the BCG some time ago and subsequently displaced from the galaxy. The offset is, however, comparable to or smaller than the size of the radio lobes, so projection effects combined with large and, possibly, asymmetric radio lobes could also cause the apparent shift. 

To traverse a distance of $22\,{\rm kpc}$ over timescales estimated above, a velocity in the range $400-5000\,{\rm km\,s^{-1}}$ is needed\footnote{This is a lower limit since we used the projected distance.}. This range overlaps with the gas velocities expected for the merger in a $10\,{\rm keV}$ cluster. Interestingly, there is another example of an `orphan' radio source in the nearby ($z=0.11$) cluster Abell 2034 \citep[source "Da", see Figure~7 in][]{2016MNRAS.459..277S}, which is some 200-300~kpc away from the BCG. Abell~2034 is also a prominent merger \citep[see, e.g.,][for recent XRISM results]{2026arXiv260427161H}, although the merger axis is closer to the sky plane than in the Peanut cluster.  The spectral index of this orphan source between 150 and 600~MHz is $\alpha\sim -1.6$ \citep[see Figure~5 in][]{2016MNRAS.459..277S} and is flatter than in Peanut, but in the ``single age''  scenario with the initial spectral index $\alpha=-0.5$, it corresponds to approximately the same age as the double-lobe source in the Peanut cluster.  

Another potentially relevant example is the radio source in the core of A2199, where the presence of an almost parallel group of radio-bright ``bridges'' connecting two radio-lobes is found \citep[][]{1983ApJ...271..575B}. One of these ``bridges'' goes through the nucleus of the galaxy, while another does not. One possible interpretation \citep[e.g.][see also \citealt{Timmerman2026} for a recent analysis and discussion]{2013ApJ...775..117N} is that gas motions (sloshing) displace the gas. However, the AGN remains active and re-establishes a new ``bridge''. The displacement in A2199 is $\sim 5\,\rm kpc$ (in projection), i.e., a factor of 4-5 smaller than in the Peanut cluster. It is therefore possible that the more extreme merger in the Peanut cluster quenched the AGN activity entirely.

The strong spectral curvature and absence of a detected compact radio core suggest that the AGN responsible for the double-lobed emission is no longer actively supplying relativistic particles to the lobes. The projected offset from G1, a possible host galaxy, may reflect displacement of the fossil plasma during the merger, although projection effects and intrinsic lobe asymmetry cannot be excluded.  If true, these are ``aged'' radio lobes.

\subsection{Displacement with the gas or a shock-bubble interaction}
\label{s:displacement}

The double-lobe source appears to be mostly co-spatial with the northern peak of the X-ray emission (see Figures~\ref{fig::optical_X-ray} and \ref{fig:shocks}), which is itself shifted to the south of the galaxy  G1. One can consider two different scenarios for the origin of this configuration. In the first scenario, the central dense core is shifted as a whole from the bottom of the potential well, centered at G1. In this case, the morphology of the radio source is modified insofar as the ISM distribution is modified during the displacement process by the ram pressure gradient acting against the restoring gravitational force.  

In the second scenario, the current shape of the double-lobe source is the result of a shock-bubble interaction, which can transform the bubble (or amorphous radio source) into a toroidal structure. The condition for this transformation is the difference in density inside and outside the bubble. See, e.g., \cite{1987JFM...181...41H,2011AnRFM..43..117R} for experimental examples of vortex-like structures arising due to the interaction of shocks with bubbles and
\cite{2002MNRAS.331.1011E,2005ApJ...634L.141H} for numerical simulations pertinent to galaxy cluster conditions. 

Higher-resolution and deeper observations are needed to differentiate between these two scenarios. For instance, for typical double-lobe sources, there is often a spectral index gradient across the structures, which, if detected, would support the first scenario. A uniform spectral index, or a steeper spectral index in the central part, would be more consistent with the second scenario, in which the shock may re-accelerate particles. At present, we consider this question open.

\subsection{Central radio halo emission}
\label{s:halo}

Another spatially and spectrally distinct radio component of the Peanut cluster is its extended diffuse emission, characterized by an ultra-steep integrated spectral index of $\alpha=-1.77\pm0.11$ between 400~MHz and 1.28~GHz (see Figures~\ref{fig::15arcsecmaps} and \ref{fig::spec}). The emission has a projected extent of approximately 1~Mpc at 400~MHz, and its radio morphology broadly follows the X-ray emission from the ICM. Radio halos are diffuse, cluster-scale synchrotron sources observed in the central regions of galaxy clusters, with morphologies that broadly trace the X-ray emitting ICM. Their emission is commonly attributed to the turbulent re-acceleration of relativistic electrons during cluster mergers \citep{Brunetti2014,vanWeeren2019}. Based on its morphology, ultra-steep spectral index, radio power, radial surface brightness profile, emissivity, and near-linear radio--X-ray surface brightness correlation, we classify the central diffuse emission in the Peanut cluster as a radio halo \citep[e.g.,][]{Pearce2017, Rajpurohit2021b, Bonafede2022, Rajpurohit2025, Santra2026}.

Ultra-steep spectrum radio halos have been observed in several merging galaxy clusters, including Abell~521 \citep{Brunetti2008, Macario2013, Santra2024}, Abell~2256 \citep{Rajpurohit2023, Osinga2024}, and MACS~J0717.5+3745 \citep{vanWeeren2017b, Rajpurohit2021b}. Their steep spectra are of particular interest in the context of turbulent re-acceleration models, which predict a population of radio halos with steep or curved spectra when particle acceleration is insufficient to compensate for radiative losses \citep[e.g.,][]{BrunettiLazarian2007, Brunetti2008, Brunetti2014}. Numerical simulations by \citet{Donnert2013} show that the radio luminosity, spatial distribution, and spectral properties of radio halos evolve during cluster mergers as turbulent re-acceleration develops and subsequently decays. In their simulations, the radio emission continues to increase after core passage as turbulence develops throughout a larger fraction of the ICM, while ultra-steep spectra arise during the later stages as particle re-acceleration becomes less efficient. 

The Peanut cluster, observed about 0.1~Gyr after pericentric passage, hosts an extended halo with an ultra-steep integrated spectrum ($\alpha_{\rm 400~MHz}^{\rm 1.28~GHz}=-1.77\pm0.11$). The presence of such a steep spectrum shortly after pericentric passage is particularly interesting, as it differs from the late-time spectral steepening found in the simulations of \citet{Donnert2013}. However, the spectral evolution of radio halos depends on several factors, for example the efficiency of turbulent re-acceleration, the magnetic field, and radiative losses. At the redshift of the Peanut cluster, enhanced IC losses against the cosmic microwave background can reduce the maximum energy attained by relativistic electrons in turbulent re-acceleration models, potentially contributing to the observed ultra-steep integrated spectrum of the halo \citep{Cassano2010, Cassano2010a}. However, the integrated spectrum of the Peanut cluster does not exhibit a high-frequency steepening. 

The Peanut halo also exhibits spatial variations in spectral index (Figure~\ref{fig::spec}, right panel). Between 815 and 1280~MHz, the spectral index ranges from approximately $-0.7$ to $-2.5$. Following the procedure described in Sect.~\ref{s:compact}, we estimate a cooling time of approximately 35~Myr for a spectrum steepening from an initial spectral index of $\alpha=-0.5$ to $-1.4$, assuming a magnetic field strength of $3.8\,\mu{\rm G}$. This cooling time is short given the observed extent of the halo. Together with the presence of patches exhibiting relatively flat spectral indices (as flat as $-0.6$), this supports the need for local particle acceleration or re-acceleration, unless they can be transported across the halo at velocities exceeding $10^{4}\,{\rm km\,s^{-1}}$ \citep[see][for a discussion of this possibility]{2026A&A...705A.184C}. 

The radial surface brightness profile of the halo (Figure~\ref{fig::profile}) is well described by a single exponential model, consistent with the profiles reported for other massive merging clusters \citep{Murgia2009, Bonafede2022, Rajpurohit2025, Santra2026}. Over the same radial range, the X-ray surface brightness exhibits a comparable decline, consistent with a linear correlation. With an e-folding radius of $r_{e}=93\pm2$~kpc, the halo lies toward the compact end of the sample studied by \citet{Murgia2009}, whose e-folding radii span approximately 70--400~kpc. Its relatively small e-folding radius and high central surface brightness yield a volume-averaged radio emissivity of $\langle J\rangle=(1.1\pm0.1)\times10^{-41}\,{\rm erg\,s^{-1}\,cm^{-3}\,Hz^{-1}}$ at 1.4~GHz (Sect.~\ref{s:spec}), approximately an order of magnitude higher than the characteristic emissivity of $\sim10^{-42}\,{\rm erg\,s^{-1}\,cm^{-3}\,Hz^{-1}}$ reported for other known halos \citep{Murgia2009, Rajpurohit2025}. Despite its high emissivity, the Peanut halo radio power is consistent with the established radio power--cluster mass scaling relation \citep[e.g.,][]{Cuciti2021, Duchesne2024}.

\subsection{Comparison with the Coma and Perseus clusters and the quenched AGN}
\label{s:peanut_against_all}
\begin{figure}[t]
    \centering
    \includegraphics[trim=0.5cm 5.0cm 1cm 2.5cm,clip,width=0.49\textwidth]{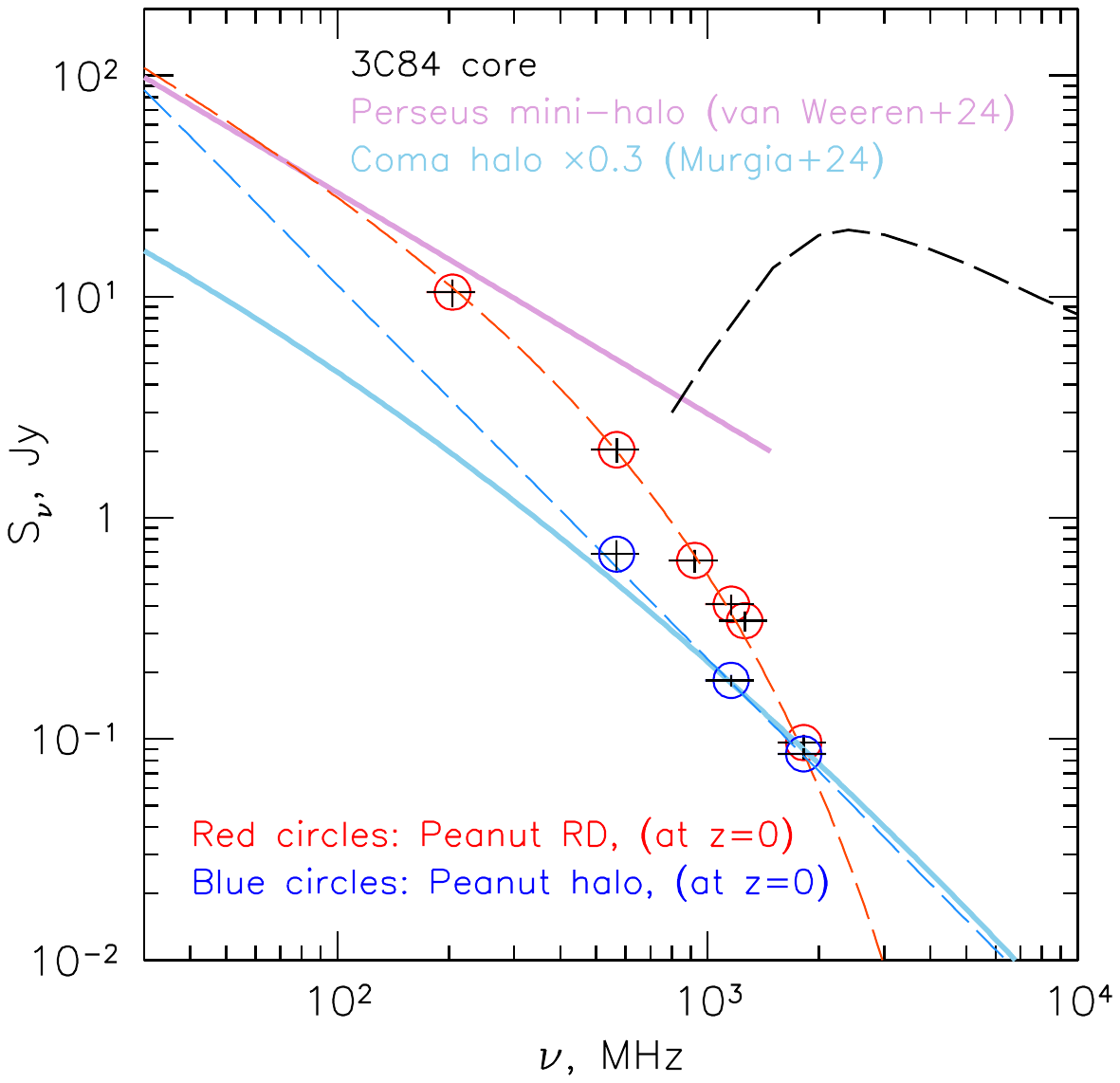}
    \caption{Comparison of the Peanut's radio components spectra (the double-lobed structure [red] and the halo [blue]) with potential counterparts in nearby clusters. The Peanut spectra were rescaled to the distance $\sim 100\,\rm Mpc$. The dashed blue line shows a power law with the spectral index of -1.77. The dashed red line shows the aged synchrotron spectrum with the initial spectral index of $-0.9$. The thick sky-blue line shows the spectrum of the Coma cluster halo \citep{2024MNRAS.528.6470M} scaled down by a factor of 3. The thick plum line shows the spectrum of the Perseus mini-halo \citep{2024A&A...692A..12V}. Finally, the black dashed line schematically shows the (time-variable) spectrum of the 3C84 nucleus. This plot illustrates the plausible correspondence between the structures found in the Peanut cluster and its low-redshift analogs. Unlike the nearby comparison systems, the Peanut cluster lacks a similarly bright compact radio source. The strong spectral curvature of its double-lobed component is consistent with an aged electron population and may constrain the time elapsed since the cessation of particle injection. }
    \label{fig::peanut-spectrum-versus-coma-perseus}
\end{figure}

The Peanut and Bullet clusters exhibit similarities in their merger morphology, although they represent different stages of merger evolution. The Peanut cluster is considered to occupy an intermediate stage between the pre-merger system MACS~J0416.1$-$2403 \citep{Ogrean2015} and the post-merger Bullet cluster \citep{2004ApJ...606..819M}. In the Peanut cluster, the galaxies associated with the northern subcluster lie slightly ahead of the bright X-ray clump, resembling the spatial offset observed in the Bullet cluster, where the X-ray gas trails the subcluster galaxies \citep{Lyskova2025}. The radio halos in the two clusters also occupy similar positions in the $I_{0}-r_{e}$ plane (Sect.~\ref{s:spec}), although they differ substantially in their spatial extent and spectral properties. The Bullet cluster hosts a considerably larger radio halo, with an e-folding radius of $r_{e}=230\pm4$~kpc and diffuse emission extending over more than 2~Mpc \citep{Sikhosana2023, Rajpurohit2025}, compared with $r_{e}=93\pm2$~kpc and an LLS of 800~kpc for the Peanut cluster. The Bullet cluster halo also has a flatter integrated radio spectrum, with $\alpha=-1.1$ \citep{Rajpurohit2025}, compared with $\alpha=-1.77\pm0.11$ for the Peanut cluster. Despite its smaller extent and steeper integrated spectrum, the Peanut halo has a high radio emissivity. Although the spectral index of the MACS~J0416.1$-$2403 halo is poorly constrained, it is consistent with that of the Peanut cluster halo, suggesting that ultra-steep spectrum halos can be formed at different stages of merger evolution.

We used the available X-ray and tSZ observations to investigate possible mechanisms for the local acceleration of relativistic electrons in the radio halo. Given the large physical extent of the halo and the short radiative lifetimes of the emitting electrons, we consider two mechanisms: particle (re)-acceleration at shocks propagating through the ICM \citep[e.g.,][]{vanWeeren2017a, Bonafede2014} and turbulent re-acceleration models \citep[e.g.,][]{Brunetti2014, Brunetti2008}. Both mechanisms may operate in merging galaxy clusters. In the shock scenario, shock related discontinuities may be detectable in the X-ray and tSZ observations and could be spatially associated with regions of flatter radio spectral index or surface brightness edges in the radio halo \citep{Botteon2023}.

Candidate surface brightness discontinuities are identified in the \textit{Chandra} residual image (Figure~\ref{fig:shocks}). The inner feature southeast of the cluster center is also present in the MUSTANG-2 residual map and spatially coincides with a region of flatter radio spectral index (Figure~\ref{fig::spec}). If confirmed as a shock, this spatial correspondence would support localized shock acceleration or re-acceleration of relativistic electrons. However, the current data do not permit reliable measurements of the gas density and temperature jumps needed to confirm the presence of a shock.

The spatial distribution and orientation of the candidate discontinuities provide additional information about their possible association with the merger. While the main merger axis is approximately north--south, the outer arc-like features are located predominantly to the east and west of the cluster center. If these features are shocks, their orientation approximately transverse to the merger axis makes an association with the primary bow shocks of the two merging components less likely. The candidate discontinuity along the northern boundary of the northern X-ray component may be more directly associated with the main merger. The gas in this component is hotter than that in the central region, although the temperature uncertainties are large \citep{Lyskova2025}, and the temperature on the opposite side of the candidate discontinuity remains unconstrained. Deeper X-ray observations are required to establish the nature of these features and assess their possible association with the radio halo emission.

It is interesting to compare the radio properties of the Peanut cluster with its local ($z\sim 0$) counterparts. This is shown in Figure~\ref{fig::peanut-spectrum-versus-coma-perseus}, which schematically illustrates the spectra of three eponymous radio sources: a radio halo (Coma-C), a mini-halo (Perseus), and a bright core of an FR-I radio source (3C84 in the Perseus core). For easier comparison, we rescaled the Peanut spectra to $z=0$ and the luminosity distance of $100\,\rm Mpc$. 

The Peanut does have an extended diffuse radio source (Sect.~\ref{s:halo}) that broadly matches the size and brightness of the Coma cluster halo (within a factor of $\sim 3$). The Peanut also has a more compact double-lobe source (Sect.~\ref{s:compact}) which, in terms of the physical size and the spectral properties, resembles Perseus's mini-halo (albeit with a steeper spectrum at the highest observed frequencies). In both cases, the explanations given above are similar to those invoked for local clusters. What is missing in the Peanut cluster is a bright FR-I (or FR-II) source similar to the nucleus of 3C84.  

As in Sect.~\ref{s:compact}, we interpret this deficiency in the same ``soon-after-the-pericenter-passage'' scenario. One possible interpretation is that the merger disrupted the gas supply to the central supermassive black hole, potentially suppressing AGN activity. If  G1 hosted the AGN responsible for the double-lobed emission, the relative motion between the galaxy and the surrounding gas may have contributed to this process. To constrain the characteristic timescale and physical parameters of this scenario, we adopt the following simplifying assumptions.

As before, we assume that the curvature of the ``radio double'' spectrum is set by the time $t_{0}$ elapsed since the AGN was quenched and the characteristic value of the magnetic field $B_0$. We further assume that the geometric offset between the galaxy and the centroid of the radio double (Figure~\ref{fig::optical_X-ray}), together with the measured LOS velocity differences between the two subclusters $\varv_{\rm LOS}$ and the angle $\theta_0$ between the merger axis and LOS, can also be used to infer $t_0$. 

This leads to the following set of constraints
\begin{align}
 &t_0= C\nu_{\rm br}^{-1/2} \frac{B_0^{1/2}}{B_0^2+B^2_{\rm CMB}} \\
 &t_0=l_{\rm SKY}/ \varv_{\rm SKY} \\
&\varv_{\rm SKY}=\varv_{\rm 3D} \sin \theta,
\end{align}
where $\nu_{\rm br}$ and $l_{\rm SKY}$ are known. From the assumption that the merger is close to the pericenter passage, one can estimate, albeit with large uncertainties, $\theta\sim \pi/4$ and the maximum relative velocity of $\varv_{\rm 3D}\sim 3500\,\rm km\,s^{-1}$ \citep{Lyskova2025}. Adopting these figures yields $t_0\sim 10\,\rm Myr$.  This is a factor of $\sim 6$ shorter than the age estimated for $B=B_{\rm CMB}/\sqrt{3}$ (see Sect.~\ref{s:compact}). Assuming that adiabatic expansion does not strongly affect the age estimate, the magnetic field $B_0\sim 10\,\rm\mu G$ is required. It falls between the two values (3.8 and 70 $\rm \mu G$) considered in Sect.~\ref{s:compact}. However, given the uncertainties in the above assumptions, e.g., the AGN might be quenched earlier, or it might remain active and connected to the lobes after the lobes are displaced, the wide range of $B_0$ (both lower and higher than $B_0\sim 10\,\rm \mu G$)  might be present in the double-lobe.

\section{Summary and conclusions}
\label{sec::summary}

We presented a multi-frequency radio study of the Peanut cluster using new MeerKAT and uGMRT observations, complemented by archival LOFAR and \textit{Chandra} data and new MUSTANG-2 observations. Our main findings are summarized below.

\begin{enumerate}
\setlength{\itemsep}{0.3em}

\item In the new observations, the brightest radio source in the cluster region is resolved into a double-lobed structure with a projected size of $\approx190$~kpc. The absence of a spectral index gradient along the adjoining ridge indicates that the source is not a head-tail radio galaxy. Its spectrum is steep and curved, suggesting that the radio plasma is no longer being actively powered by an AGN. The radio centroid of this source is offset by $\approx22$~kpc from the nearest brightest cluster galaxy. We suggest that the source represents aged fossil plasma that may have been displaced or reshaped during the merger. Both the displacement of fossil radio lobes and the shock-induced reshaping of an older radio bubble remain possible explanations for its observed morphology.

\item We report the first detection of extended diffuse radio emission with an LLS of $\approx1$~Mpc, which we identify as an ultra-steep spectrum halo. Its brightest part, the ridge, extends over $\approx400$~kpc southward from the northern subcluster and closely follows the chain-like distribution of optical galaxies. The halo has an integrated spectral index of $\alpha_{\rm 400~MHz}^{\rm 1.28~GHz}=-1.77\pm0.11$.  An ultra-steep spectrum halo at such an early merger stage is unusual, since in the turbulent re-acceleration framework steep spectra are generally associated with the later merger stages. The spatially resolved spectral index distribution reveals variations across the halo. Given the short radiative lifetimes of relativistic electrons, these variations suggest ongoing acceleration or re-acceleration of particles.

\item The radial surface brightness profile of the halo is well described by a single exponential model with an e-folding radius of $r_{e}=93\pm2$~kpc. Its volume-averaged radio emissivity is $(1.1\pm0.1)\times10^{-41}\,{\rm erg\,s^{-1}\,cm^{-3}\,Hz^{-1}}$ at 1.4~GHz, approximately an order of magnitude higher than the characteristic value reported for other halos but its radio power is consistent with the known scaling relations for halos. The Peanut halo is smaller and has a steeper integrated spectrum and higher radio emissivity than the halo in the Bullet cluster.

\item The radio and X-ray profiles exhibit a similar decline with increasing distance from the cluster center. Also, the radio and X-ray emission in the halo region is tightly correlated, with a best-fit slope of $1.11\pm0.05$, indicating a close coupling between the thermal and nonthermal emission.

\item The residual \textit{Chandra} X-ray and thermal SZ maps reveal three arc-like candidate surface brightness discontinuities, with the most prominent inner feature spatially coincident with a region of spectral flattening. If this discontinuity is associated with a shock, the coincident spectral flattening may indicate localized particle acceleration or re-acceleration. The nature of these discontinuities cannot be established with the current data.

\item Comparison with nearby clusters indicates that the Peanut cluster contains radio components broadly analogous to the Coma radio halo and the Perseus mini-halo, but lacks a bright compact radio source similar to 3C~84 in the Perseus cluster. This supports the interpretation that the observed double-lobe structure is a remnant of recently quenched AGN activity associated with the ongoing cluster merger.
\end{enumerate}

Our results underline the importance of deep radio observations for detecting and characterizing faint diffuse radio emission associated with the ICM. In the Peanut cluster, although the radio halo is recovered at frequencies above 400~MHz, its non-detection in the available LOFAR HBA data (owing to poor data quality) prevents us from fully characterizing its low-frequency spectral properties and true spatial extent. Deeper low-frequency observations with LOFAR2.0 will be crucial for probing the aged relativistic electron population, determining the broadband spectral shape of the radio halo, and constraining models of particle acceleration. At the same time, subarcsecond radio observations, for example with LOFAR VLBI, and high-frequency polarization observations will be essential for resolving the internal structure of the compact double-lobe source and distinguishing between merger-induced displacement of fossil AGN plasma and shock-driven reshaping of an aged radio bubble.

Finally, deeper X-ray observations, complemented by high-resolution spectroscopic measurements from XRISM and future thermal and kinematic SZ studies, will be required to characterize the candidate surface brightness discontinuities and determine whether they are associated with merger shocks. Such observations could help constrain the role of shocks and turbulence in the acceleration of relativistic particles and clarify the connection between the thermal ICM, fossil AGN plasma, and the origin and evolution of the ultra-steep spectrum halo during a major cluster merger. The Peanut cluster therefore provides an important test case for investigating the connection between the thermal and dynamical evolution of the ICM and the origin of ultra-steep spectrum radio halo emission with high radio emissivity shortly after pericentric passage.

\section*{Acknowledgements}
WF, CJ, and RK acknowledge support from the Smithsonian Institution and the \textit{Chandra} High Resolution Camera Project through NASA contract NAS8-03060. 
TM acknowledges support from the Agencia Estatal de Investigaci\'on (AEI) and the Ministerio de Ciencia, Innovaci\'on y Universidades (MICIU) Grant ATRAE2024-154740 funded by MICIU/AEI//10.13039/501100011033.
TM is also partly supported by the Spanish program Unidad de Excelencia Mar\'{\i}a de Maeztu CEX2020-001058-M, financed by MCIN/AEI//10.13039/501100011033, and by the MaX-CSIC Excellence Award MaX4-SOMMA-ICE. IK was supported by the Simons Foundation via the Simons Investigator Award to A. A. Schekochihin. CZ acknowledges the support of the Czech Science Foundation (GACR) Junior Star grant no. GM24-10599M.

The authors thank the staff of the MeerKAT Observatory for their help with the observations presented in this work. The MeerKAT telescope is operated by the South African Radio Astronomy Observatory, which is a facility of the National Research Foundation, an agency of the Department of Science and Innovation. We thank the staff of the GMRT that made these observations possible. GMRT is run by the National Centre for Radio Astrophysics of the Tata Institute of Fundamental Research. LOFAR \citep{Haarlem2013} is the Low Frequency Array designed and constructed by ASTRON. It has observing, data processing, and data storage facilities in several countries, which are owned by various parties (each with their own funding sources), and that are collectively operated by the ILT foundation under a joint scientific policy. The radio data reduction was conducted on the Smithsonian High Performance Cluster (SI/HPC), Smithsonian Institution (\url{https://doi.org/10.25572/SIHPC}).  And finally, the authors thank the staff at the Green Bank Observatory for support in operating and maintaining the MUSTANG-2 instrument on the Green Bank Telescope.
MUSTANG-2 is supported by the U.S.\ National Science Foundation (NSF) award number 1615604.
The National Radio Astronomy Observatory and Green Bank Observatory are facilities of the NSF operated under cooperative agreement by Associated Universities, Inc. 
The GBT data were acquired under the project ID AGBT25B\_215. 

\bibliographystyle{aa} 
\bibliography{ref}

@ARTICLE{Govoni2001,
       author = {{Govoni}, F. and {Enßlin}, T.~A. and {Feretti}, L. and {Giovannini}, G.},
        title = "{A comparison of radio and X-ray morphologies of four clusters of galaxies containing radio halos}",
      journal = {\aap},
         year = 2001,
       volume = {369},
        pages = {441-449},
          doi = {10.1051/0004-6361:20010115},
       adsurl = {https://ui.adsabs.harvard.edu/abs/2001A&A...369..441G}
}

@ARTICLE{Murgia2009,
       author = {{Murgia}, M. and {Govoni}, F. and {Markevitch}, M. and {Feretti}, L. and {Giovannini}, G. and {Taylor}, G.~B. and {Carretti}, E.},
        title = "{Comparative analysis of the diffuse radio emission in the galaxy clusters A1835, A2029, A2142, A2255, A2319, A3562 and 1E 0657-56}",
      journal = {\aap},
         year = 2009,
        month = jun,
       volume = {499},
       number = {3},
        pages = {679-695},
          doi = {10.1051/0004-6361/200811205},
archivePrefix = {arXiv},
       eprint = {0901.1943},
 primaryClass = {astro-ph.CO},
       adsurl = {https://ui.adsabs.harvard.edu/abs/2009A&A...499..679M}
}

@ARTICLE{Sikhosana2023,
       author = {{Sikhosana}, S.~P. and {Knowles}, K. and {Hilton}, M. and {Moodley}, K. and {Murgia}, M.},
        title = "{MeerKAT's view of the Bullet cluster 1E 0657-55.8}",
      journal = {\mnras},
         year = 2023,
       volume = {518},
        pages = {4595},
          doi = {10.1093/mnras/stac3370},
archivePrefix = {arXiv},
       eprint = {2207.05492},
 primaryClass = {astro-ph.CO},
       adsurl = {https://ui.adsabs.harvard.edu/abs/2023MNRAS.518.4595S}
}

@ARTICLE{Brienza2022,
       author = {{Brienza}, M. and {Lovisari}, L. and {Rajpurohit}, K. and {Bonafede}, A. and {Gastaldello}, F. and {Murgia}, M. and {Vazza}, F. and {Bonnassieux}, E. and {Botteon}, A. and {Brunetti}, G. and {Drabent}, A. and {Hardcastle}, M.~J. and {Pasini}, T. and {Riseley}, C.~J. and {R{\"o}ttgering}, H.~J.~A. and {Shimwell}, T. and {Simionescu}, A. and {van Weeren}, R.~J.},
        title = "{The galaxy group NGC 507: Newly detected AGN remnant plasma transported by sloshing}",
      journal = {\aap},
         year = 2022,
        month = may,
       volume = {661},
          eid = {A92},
        pages = {A92},
          doi = {10.1051/0004-6361/202142579},
archivePrefix = {arXiv},
       eprint = {2201.04591},
 primaryClass = {astro-ph.GA},
       adsurl = {https://ui.adsabs.harvard.edu/abs/2022A&A...661A..92B}
}

@ARTICLE{Mandal2024,
       author = {{Mandal}, S. and {Intema}, H.~T. and {van Weeren}, R.~J. and {Shimwell}, T.~W. and {Botteon}, A. and {Brunetti}, G. and {de Gasperin}, F. and {Br{\"u}ggen}, M. and {Di Gennaro}, G. and {Kraft}, R. and {R{\"o}ttgering}, H.~J.~A. and {Hardcastle}, M. and {Tasse}, C.},
        title = "{Revived fossil plasma sources in galaxy clusters}",
      journal = {\aap},
         year = 2020,
        month = feb,
       volume = {634},
          eid = {A4},
        pages = {A4},
          doi = {10.1051/0004-6361/201936560},
archivePrefix = {arXiv},
       eprint = {1911.02034},
 primaryClass = {astro-ph.CO},
       adsurl = {https://ui.adsabs.harvard.edu/abs/2020A&A...634A...4M}
}

@ARTICLE{Rajpurohit2024a,
       author = {{Rajpurohit}, K. and {O'Sullivan}, E. and {Schellenberger}, G. and {Brienza}, M. and {Vrtilek}, J.~M. and {Forman}, W. and {David}, L.~P. and {Clarke}, T. and {Botteon}, A. and {Vazza}, F. and {Giacintucci}, S. and {Jones}, C. and {Br{\"u}ggen}, M. and {Shimwell}, T.~W. and {Drabent}, A. and {Loi}, F. and {Loubser}, S.~I. and {Kolokythas}, K. and {Babyk}, I. and {R{\"o}ttgering}, H.~J.~A.},
        title = "{A Deep Dive into the NGC 741 Galaxy Group: Insights into a Spectacular Head-tail Radio Galaxy from VLA, MeerKAT, uGMRT, and LOFAR}",
      journal = {\apj},
         year = 2024,
        month = nov,
       volume = {976},
       number = {1},
          eid = {64},
        pages = {64},
          doi = {10.3847/1538-4357/ad8136},
archivePrefix = {arXiv},
       eprint = {2408.15197},
 primaryClass = {astro-ph.GA},
       adsurl = {https://ui.adsabs.harvard.edu/abs/2024ApJ...976...64R}
}

@ARTICLE{JaffePerola1973,
       author = {{Jaffe}, W.~J. and {Perola}, G.~C.},
        title = "{Dynamical Models of Tailed Radio Sources in Clusters of Galaxies}",
      journal = {\aap},
         year = 1973,
        month = aug,
       volume = {26},
        pages = {423},
       adsurl = {https://ui.adsabs.harvard.edu/abs/1973A&A....26..423J}
}

@ARTICLE{Botteon2026b,
       author = {{Botteon}, A. and {Brienza}, M. and {Rajpurohit}, K. and {Lyskova}, N. and {Churazov}, E. and {Khabibullin}, I. and {Pasini}, T. and {O'Sullivan}, E. and {Brunetti}, G. and {De Gasperin}, F. and {De Rubeis}, E. and {Gastaldello}, F. and {Hoang}, D.~N. and {Kraft}, R. and {Schellenberger}, G. and {Sunyaev}, R. and {van Weeren}, R.~J. and {Vazza}, F.},
        title = "{The Treble Clef radio phoenix and its old nonthermal filaments}",
      journal = {arXiv e-prints},
         year = 2026,
       eprint = {2607.06659},
 primaryClass = {astro-ph.CO},
}

@ARTICLE{Santra2026,
       author = {{Santra}, R. and {Balboni}, M. and {Kale}, R.},
        title = "{Multi-wavelength signatures of a giant cometary radio halo in MACS J0417-1154}",
      journal = {arXiv e-prints},
         year = 2026,
       eprint = {2607.07750},
 primaryClass = {astro-ph.CO},
}

@ARTICLE{Timmerman2026,
       author = {{Timmerman}, R. and {Rudnick}, L. and {Botteon}, A. and {Brunetti}, G. and {Kale}, R.},
        title = "{The magnetic mayhem in Abell 2199: discovery of synchrotron threads and homogeneous diffuse radio lobes}",
      journal = {\mnras},
         year = 2026,
          doi = {10.1093/mnras/stag1232},
       eprint = {2606.30714},
}

@ARTICLE{Kelly2007,
       author = {{Kelly}, Brandon C.},
        title = "{Some Aspects of Measurement Error in Linear Regression of Astronomical Data}",
      journal = {\apj},
         year = 2007,
        month = aug,
       volume = {665},
       number = {2},
        pages = {1489-1506},
          doi = {10.1086/519947},
       adsurl = {https://ui.adsabs.harvard.edu/abs/2007ApJ...665.1489K},
}

@ARTICLE{deGasperin2019,
       author = {{de Gasperin}, F. and {Dijkema}, T.~J. and {Drabent}, A. and {Mevius}, M. and {Rafferty}, D. and {van Weeren}, R. and {Br{\"u}ggen}, M. and {Callingham}, J.~R. and {Emig}, K.~L. and {Heald}, G. and {Intema}, H.~T. and {Morabito}, L.~K. and {Offringa}, A.~R. and {Oonk}, R. and {Orr{\`u}}, E. and {R{\"o}ttgering}, H. and {Sabater}, J. and {Shimwell}, T. and {Shulevski}, A. and {Williams}, W.},
        title = "{Systematic effects in LOFAR data: A unified calibration strategy}",
      journal = {\aap},
         year = 2019,
        month = feb,
       volume = {622},
          eid = {A5},
        pages = {A5},
          doi = {10.1051/0004-6361/201833867},
archivePrefix = {arXiv},
       eprint = {1811.07954},
 primaryClass = {astro-ph.IM},
       adsurl = {https://ui.adsabs.harvard.edu/abs/2019A&A...622A...5D}
}

@ARTICLE{vanWeeren2016a,
       author = {{van Weeren}, R.~J. and {Williams}, W.~L. and {Hardcastle}, M.~J. and {Shimwell}, T.~W. and {Rafferty}, D.~A. and {Sabater}, J. and {Heald}, G. and {Sridhar}, S.~S. and {Dijkema}, T.~J. and {Brunetti}, G. and {Br{\"u}ggen}, M. and {Andrade-Santos}, F. and {Ogrean}, G.~A. and {R{\"o}ttgering}, H.~J.~A. and {Dawson}, W.~A. and {Forman}, W.~R. and {de Gasperin}, F. and {Jones}, C. and {Miley}, G.~K. and {Rudnick}, L. and {Sarazin}, C.~L. and {Bonafede}, A. and {Best}, P.~N. and {B{\^\i}rzan}, L. and {Cassano}, R. and {Chy{\.z}y}, K.~T. and {Croston}, J.~H. and {Ensslin}, T. and {Ferrari}, C. and {Hoeft}, M. and {Horellou}, C. and {Jarvis}, M.~J. and {Kraft}, R.~P. and {Mevius}, M. and {Intema}, H.~T. and {Murray}, S.~S. and {Orr{\'u}}, E. and {Pizzo}, R. and {Simionescu}, A. and {Stroe}, A. and {van der Tol}, S. and {White}, G.~J.},
        title = "{LOFAR Facet Calibration}",
      journal = {\apjs},
         year = 2016,
        month = mar,
       volume = {223},
       number = {1},
          eid = {2},
        pages = {2},
          doi = {10.3847/0067-0049/223/1/2},
archivePrefix = {arXiv},
       eprint = {1601.05422},
 primaryClass = {astro-ph.IM},
       adsurl = {https://ui.adsabs.harvard.edu/abs/2016ApJS..223....2V}
}

@ARTICLE{Williams2016,
       author = {{Williams}, W.~L. and {van Weeren}, R.~J. and {R{\"o}ttgering}, H.~J.~A. and {Best}, P. and {Dijkema}, T.~J. and {de Gasperin}, F. and {Hardcastle}, M.~J. and {Heald}, G. and {Prandoni}, I. and {Sabater}, J. and {Shimwell}, T.~W. and {Tasse}, C. and {van Bemmel}, I.~M. and {Br{\"u}ggen}, M. and {Brunetti}, G. and {Conway}, J.~E. and {En{\ss}lin}, T. and {Engels}, D. and {Falcke}, H. and {Ferrari}, C. and {Haverkorn}, M. and {Jackson}, N. and {Jarvis}, M.~J. and {Kapi{\'n}ska}, A.~D. and {Mahony}, E.~K. and {Miley}, G.~K. and {Morabito}, L.~K. and {Morganti}, R. and {Orr{\'u}}, E. and {Retana-Montenegro}, E. and {Sridhar}, S.~S. and {Toribio}, M.~C. and {White}, G.~J. and {Wise}, M.~W. and {Zwart}, J.~T.~L.},
        title = "{LOFAR 150-MHz observations of the Bo{\"o}tes field: catalogue and source counts}",
      journal = {\mnras},
         year = 2016,
        month = aug,
       volume = {460},
       number = {3},
        pages = {2385-2412},
          doi = {10.1093/mnras/stw1056},
archivePrefix = {arXiv},
       eprint = {1605.01531},
 primaryClass = {astro-ph.CO},
       adsurl = {https://ui.adsabs.harvard.edu/abs/2016MNRAS.460.2385W}
}

@ARTICLE{Shimwell2026,
       author = {{Shimwell}, T.~W. and {Hardcastle}, M.~J. and {Tasse}, C. and {Drabent}, A. and {Botteon}, A. and {Williams}, W.~L. and {Best}, P.~N. and {R{\"o}ttgering}, H.~J.~A. and {Br{\"u}ggen}, M. and {Brunetti}, G. and {Callingham}, J.~R. and {Chy{\.z}y}, K.~T. and {Conway}, J.~E. and {De Gasperin}, F. and {Haverkorn}, M. and {Horellou}, C. and {Jackson}, N. and {Miley}, G.~K. and {Morabito}, L.~K. and {Morganti}, R. and {O'Sullivan}, S.~P. and {Schwarz}, D.~J. and {Smith}, D.~J.~B. and {van Weeren}, R.~J. and {Vedantham}, H.~K. and {White}, G.~J. and {Ahmadi}, A. and {Alegre}, L. and {Arias}, M. and {Asabere}, B. and {Bahr-Kalus}, B. and {Barkus}, B. and {Bilicki}, M. and {B{\"o}hme}, L. and {Brentjens}, M. and {Brienza}, M. and {Bomans}, D.~J. and {Bonafede}, A. and {Bonato}, M. and {Bonnassieux}, E. and {Boxelaar}, J.~M. and {Camera}, S. and {Cassano}, R. and {Chilufya}, J. and {Cianfaglione}, M. and {Croston}, J.~H. and {Cuciti}, V. and {Dabhade}, P. and {De Rubeis}, E. and {de Jong}, J.~M.~G.~H.~J. and {Dallacasa}, D. and {Dettmar}, R.~J. and {Duncan}, K.~J. and {Di Gennaro}, G. and {Edler}, H.~W. and {Groeneveld}, C. and {G{\"u}rkan}, G. and {Hajduk}, M. and {Hale}, C.~L. and {Heesen}, V. and {Hoang}, D.~N. and {Hoeft}, M. and {Holties}, H. and {Horton}, M.~A. and {Iacobelli}, M. and {Jamrozy}, M. and {Jarvis}, M.~J. and {Jelic}, V. and {Kadler}, M. and {Kondapally}, R. and {Kunert-Bajraszewska}, M. and {Loose}, M. and {Magliocchetti}, M. and {Ma{\l}ek}, K. and {Manzano}, C. and {McKean}, J.~P. and {Mevius}, M. and {Mingo}, B. and {Miskolczi}, A. and {Misra}, A. and {Mold{\'o}n}, J. and {Nair}, D.~G. and {Nakoneczny}, S.~J. and {Orru}, E. and {Pashapour-Ahmadabadi}, M. and {Pasini}, T. and {Petley}, J. and {Pierce}, J.~C.~S. and {Prandoni}, I. and {Rafferty}, D. and {Rajpurohit}, K. and {Riseley}, C.~J. and {Roberts}, I.~D. and {Sethi}, S. and {Shulevski}, A. and {Stein}, M. and {Stuardi}, C. and {Sweijen}, F. and {ter Veen}, S. and {Timmerman}, R. and {Vaccari}, M. and {Wijnholds}, S.},
        title = "{The LOFAR Two-metre Sky Survey: VII. Third Data Release}",
      journal = {\aap},
         year = 2026,
        month = mar,
       volume = {707},
          eid = {A198},
        pages = {A198},
          doi = {10.1051/0004-6361/202557749},
archivePrefix = {arXiv},
       eprint = {2602.15949},
 primaryClass = {astro-ph.GA},
       adsurl = {https://ui.adsabs.harvard.edu/abs/2026A&A...707A.198S}
}

@ARTICLE{Shimwell2019,
       author = {{Shimwell}, T.~W. and {Tasse}, C. and {Hardcastle}, M.~J. and {Mechev}, A.~P. and {Williams}, W.~L. and {Best}, P.~N. and {R{\"o}ttgering}, H.~J.~A. and {Callingham}, J.~R. and {Dijkema}, T.~J. and {de Gasperin}, F. and {Hoang}, D.~N. and {Hugo}, B. and {Mirmont}, M. and {Oonk}, J.~B.~R. and {Prandoni}, I. and {Rafferty}, D. and {Sabater}, J. and {Smirnov}, O. and {van Weeren}, R.~J. and {White}, G.~J. and {Atemkeng}, M. and {Bester}, L. and {Bonnassieux}, E. and {Br{\"u}ggen}, M. and {Brunetti}, G. and {Chy{\.z}y}, K.~T. and {Cochrane}, R. and {Conway}, J.~E. and {Croston}, J.~H. and {Danezi}, A. and {Duncan}, K. and {Haverkorn}, M. and {Heald}, G.~H. and {Iacobelli}, M. and {Intema}, H.~T. and {Jackson}, N. and {Jamrozy}, M. and {Jarvis}, M.~J. and {Lakhoo}, R. and {Mevius}, M. and {Miley}, G.~K. and {Morabito}, L. and {Morganti}, R. and {Nisbet}, D. and {Orr{\'u}}, E. and {Perkins}, S. and {Pizzo}, R.~F. and {Schrijvers}, C. and {Smith}, D.~J.~B. and {Vermeulen}, R. and {Wise}, M.~W. and {Alegre}, L. and {Bacon}, D.~J. and {van Bemmel}, I.~M. and {Beswick}, R.~J. and {Bonafede}, A. and {Botteon}, A. and {Bourke}, S. and {Brienza}, M. and {Calistro Rivera}, G. and {Cassano}, R. and {Clarke}, A.~O. and {Conselice}, C.~J. and {Dettmar}, R.~J. and {Drabent}, A. and {Dumba}, C. and {Emig}, K.~L. and {En{\ss}lin}, T.~A. and {Ferrari}, C. and {Garrett}, M.~A. and {G{\'e}nova-Santos}, R.~T. and {Goyal}, A. and {G{\"u}rkan}, G. and {Hale}, C. and {Harwood}, J.~J. and {Heesen}, V. and {Hoeft}, M. and {Horellou}, C. and {Jackson}, C. and {Kokotanekov}, G. and {Kondapally}, R. and {Kunert-Bajraszewska}, M. and {Mahatma}, V. and {Mahony}, E.~K. and {Mandal}, S. and {McKean}, J.~P. and {Merloni}, A. and {Mingo}, B. and {Miskolczi}, A. and {Mooney}, S. and {Nikiel-Wroczy{\'n}ski}, B. and {O'Sullivan}, S.~P. and {Quinn}, J. and {Reich}, W. and {Roskowi{\'n}ski}, C. and {Rowlinson}, A. and {Savini}, F. and {Saxena}, A. and {Schwarz}, D.~J. and {Shulevski}, A. and {Sridhar}, S.~S. and {Stacey}, H.~R. and {Urquhart}, S. and {van der Wiel}, M.~H.~D. and {Varenius}, E. and {Webster}, B. and {Wilber}, A.},
        title = "{The LOFAR Two-metre Sky Survey. II. First data release}",
      journal = {\aap},
         year = 2019,
        month = feb,
       volume = {622},
          eid = {A1},
        pages = {A1},
          doi = {10.1051/0004-6361/201833559},
archivePrefix = {arXiv},
       eprint = {1811.07926},
 primaryClass = {astro-ph.GA},
       adsurl = {https://ui.adsabs.harvard.edu/abs/2019A&A...622A...1S}
}

@ARTICLE{Botteon2026,
       author = {{Botteon}, Andrea and {Caglar}, Turgay and {D{\"o}ner}, Sibel and {van Weeren}, Reinout J. and {Smith}, Krista Lynne},
        title = "{Enhanced radio emission between a galaxy cluster pair}",
      journal = {\aap},
         year = 2026,
        month = jan,
       volume = {705},
          eid = {A230},
        pages = {A230},
          doi = {10.1051/0004-6361/202556824},
archivePrefix = {arXiv},
       eprint = {2509.14348},
 primaryClass = {astro-ph.CO},
       adsurl = {https://ui.adsabs.harvard.edu/abs/2026A&A...705A.230B}
}

@ARTICLE{Botteon2023,
       author = {{Botteon}, Andrea and {Markevitch}, Maxim and {van Weeren}, Reinout J. and {Brunetti}, Gianfranco and {Shimwell}, Timothy W.},
        title = "{Surface brightness discontinuities in radio halos. Insights from the MeerKAT Galaxy Cluster Legacy Survey}",
      journal = {\aap},
         year = 2023,
        month = jun,
       volume = {674},
          eid = {A53},
        pages = {A53},
          doi = {10.1051/0004-6361/202346150},
archivePrefix = {arXiv},
       eprint = {2302.07881},
 primaryClass = {astro-ph.CO},
       adsurl = {https://ui.adsabs.harvard.edu/abs/2023A&A...674A..53B}
}

@ARTICLE{Bonafede2014,
       author = {{Bonafede}, A. and {Intema}, H.~T. and {Br{\"u}ggen}, M. and
         {Girardi}, M. and {Nonino}, M. and {Kantharia}, N. and
         {van Weeren}, R.~J. and {R{\"o}ttgering}, H.~J.~A.},
        title = "{Evidence for Particle Re-acceleration in the Radio Relic in the Galaxy Cluster PLCKG287.0+32.9}",
      journal = {\apj},
         year = 2014,
        month = apr,
       volume = {785},
       number = {1},
          eid = {1},
        pages = {1},
          doi = {10.1088/0004-637X/785/1/1},
archivePrefix = {arXiv},
       eprint = {1402.1492},
 primaryClass = {astro-ph.CO},
       adsurl = {https://ui.adsabs.harvard.edu/abs/2014ApJ...785....1B}
}

@ARTICLE{2022MNRAS.514.1477R,
       author = {{Russell}, H.~R. and {Nulsen}, P.~E.~J. and {Caprioli}, D. and {Chadayammuri}, U. and {Fabian}, A.~C. and {Kunz}, M.~W. and {McNamara}, B.~R. and {Sanders}, J.~S. and {Richard-Laferri{\`e}re}, A. and {Beleznay}, M. and {Canning}, R.~E.~A. and {Hlavacek-Larrondo}, J. and {King}, L.~J.},
        title = "{The structure of cluster merger shocks: turbulent width and the electron heating time-scale}",
      journal = {\mnras},
         year = 2022,
        month = jul,
       volume = {514},
       number = {1},
        pages = {1477-1493},
          doi = {10.1093/mnras/stac1055},
archivePrefix = {arXiv},
       eprint = {2204.05785},
 primaryClass = {astro-ph.HE},
       adsurl = {https://ui.adsabs.harvard.edu/abs/2022MNRAS.514.1477R}
}

@ARTICLE{vanWeeren2017a,
   author = {{van Weeren}, R.~J. and {Andrade-Santos}, F. and {Dawson}, W.~A. and 
	{Golovich}, N. and {Lal}, D.~V. and {Kang}, H. and {Ryu}, D. and 
	{Br{\`i}ggen}, M. and {Ogrean}, G.~A. and {Forman}, W.~R. and 
	{Jones}, C. and {Placco}, V.~M. and {Santucci}, R.~M. and {Wittman}, D. and 
	{Jee}, M.~J. and {Kraft}, R.~P. and {Sobral}, D. and {Stroe}, A. and 
	{Fogarty}, K.},
    title = "{The case for electron re-acceleration at galaxy cluster shocks}",
  journal = {Nature Astronomy},
archivePrefix = "arXiv",
   eprint = {1701.01439},
 primaryClass = "astro-ph.HE",
     year = 2017,
    month = jan,
   volume = 1,
      eid = {0005},
    pages = {0005},
      doi = {10.1038/s41550-016-0005},
   adsurl = {http://adsabs.harvard.edu/abs/2017NatAs...1E...5V}
}

@ARTICLE{Brunetti2008,
       author = {{Brunetti}, G. and {Giacintucci}, S. and {Cassano}, R. and {Lane}, W. and {Dallacasa}, D. and {Venturi}, T. and {Kassim}, N.~E. and {Setti}, G. and {Cotton}, W.~D. and {Markevitch}, M.},
        title = "{A low-frequency radio halo associated with a cluster of galaxies}",
      journal = {\nat},
         year = 2008,
        month = oct,
       volume = {455},
       number = {7215},
        pages = {944-947},
          doi = {10.1038/nature07379},
archivePrefix = {arXiv},
       eprint = {0810.4288},
 primaryClass = {astro-ph},
       adsurl = {https://ui.adsabs.harvard.edu/abs/2008Natur.455..944B}
}

@ARTICLE{Duchesne2024,
       author = {{Duchesne}, S.~W. and {Botteon}, A. and {Koribalski}, B.~S. and {Loi}, F. and {Rajpurohit}, K. and {Riseley}, C.~J. and {Rudnick}, L. and {Vernstrom}, T. and {Andernach}, H. and {Hopkins}, A.~M. and {Kapinska}, A.~D. and {Norris}, R.~P. and {Zafar}, T.},
        title = "{Evolutionary Map of the Universe (EMU): A pilot search for diffuse, non-thermal radio emission in galaxy clusters with the Australian SKA Pathfinder}",
      journal = {\pasa},
         year = 2024,
        month = jan,
       volume = {41},
          eid = {e026},
        pages = {e026},
          doi = {10.1017/pasa.2024.10},
archivePrefix = {arXiv},
       eprint = {2402.06192},
 primaryClass = {astro-ph.CO},
       adsurl = {https://ui.adsabs.harvard.edu/abs/2024PASA...41...26D}
}

@ARTICLE{2024A&A...686A...5B,
       author = {{Balboni}, M. and {Gastaldello}, F. and {Bonafede}, A. and {Botteon}, A. and {Bartalucci}, I. and {Bourdin}, H. and {Brunetti}, G. and {Cassano}, R. and {De Grandi}, S. and {De Luca}, F. and {Ettori}, S. and {Ghizzardi}, S. and {Gitti}, M. and {Iqbal}, A. and {Johnston-Hollitt}, M. and {Lovisari}, L. and {Mazzotta}, P. and {Molendi}, S. and {Pointecouteau}, E. and {Pratt}, G.~W. and {Riva}, G. and {Rossetti}, M. and {Rottgering}, H. and {Sereno}, M. and {van Weeren}, R.~J. and {Venturi}, T. and {Veronesi}, I.},
        title = "{CHEX-MATE: A LOFAR pilot X-ray - radio study on five radio halo clusters}",
      journal = {\aap},
         year = 2024,
        month = jun,
       volume = {686},
          eid = {A5},
        pages = {A5},
          doi = {10.1051/0004-6361/202347965},
archivePrefix = {arXiv},
       eprint = {2402.18654},
 primaryClass = {astro-ph.CO},
       adsurl = {https://ui.adsabs.harvard.edu/abs/2024A&A...686A...5B}
}

@ARTICLE{Botteon2020,
       author = {{Botteon}, A. and {Brunetti}, G. and {van Weeren}, R.~J. and {Shimwell}, T.~W. and {Pizzo}, R.~F. and {Cassano}, R. and {Iacobelli}, M. and {Gastaldello}, F. and {B{\^\i}rzan}, L. and {Bonafede}, A. and {Br{\"u}ggen}, M. and {Cuciti}, V. and {Dallacasa}, D. and {de Gasperin}, F. and {Di Gennaro}, G. and {Drabent}, A. and {Hardcastle}, M.~J. and {Hoeft}, M. and {Mandal}, S. and {R{\"o}ttgering}, H.~J.~A. and {Simionescu}, A.},
        title = "{The Beautiful Mess in Abell 2255}",
      journal = {\apj},
         year = 2020,
        month = jul,
       volume = {897},
       number = {1},
          eid = {93},
        pages = {93},
          doi = {10.3847/1538-4357/ab9a2f},
archivePrefix = {arXiv},
       eprint = {2006.04808},
 primaryClass = {astro-ph.GA},
       adsurl = {https://ui.adsabs.harvard.edu/abs/2020ApJ...897...93B}
}

@ARTICLE{Riseley2024,
       author = {{Riseley}, C.~J. and {Bonafede}, A. and {Bruno}, L. and {Botteon}, A. and {Rossetti}, M. and {Biava}, N. and {Bonnassieux}, E. and {Loi}, F. and {Vernstrom}, T. and {Balboni}, M.},
        title = "{A ``MeerKAT-meets-LOFAR'' study of the complex multi-component (mini-)halo in the extreme sloshing cluster Abell 2142}",
      journal = {\aap},
         year = 2024,
        month = jun,
       volume = {686},
          eid = {A44},
        pages = {A44},
          doi = {10.1051/0004-6361/202348944},
archivePrefix = {arXiv},
       eprint = {2403.00414},
 primaryClass = {astro-ph.CO},
       adsurl = {https://ui.adsabs.harvard.edu/abs/2024A&A...686A..44R}
}

@ARTICLE{Chandra2017,
       author = {{Chandra}, Poonam and {Kanekar}, Nissim},
        title = "{Giant Metrewave Radio Telescope Monitoring of the Black Hole X-Ray Binary, V404 Cygni during Its 2015 June Outburst}",
      journal = {\apj},
         year = 2017,
        month = sep,
       volume = {846},
       number = {2},
          eid = {111},
        pages = {111},
          doi = {10.3847/1538-4357/aa85a2},
archivePrefix = {arXiv},
       eprint = {1708.02739},
 primaryClass = {astro-ph.HE},
       adsurl = {https://ui.adsabs.harvard.edu/abs/2017ApJ...846..111C}
}

@ARTICLE{McConnell2020,
       author = {{McConnell}, D. and {Hale}, C.~L. and {Lenc}, E. and {Banfield}, J.~K. and {Heald}, George and {Hotan}, A.~W. and {Leung}, James K. and {Moss}, Vanessa A. and {Murphy}, Tara and {O'Brien}, Andrew and {Pritchard}, Joshua and {Raja}, Wasim and {Sadler}, Elaine M. and {Stewart}, Adam and {Thomson}, Alec J.~M. and {Whiting}, M. and {Allison}, James R. and {Amy}, S.~W. and {Anderson}, C. and {Ball}, Lewis and {Bannister}, Keith W. and {Bell}, Martin and {Bock}, Douglas C.-J. and {Bolton}, Russ and {Bunton}, J.~D. and {Chippendale}, A.~P. and {Collier}, J.~D. and {Cooray}, F.~R. and {Cornwell}, T.~J. and {Diamond}, P.~J. and {Edwards}, P.~G. and {Gupta}, N. and {Hayman}, Douglas B. and {Heywood}, Ian and {Jackson}, C.~A. and {Koribalski}, B{\"a}rbel S. and {Lee-Waddell}, Karen and {McClure-Griffiths}, N.~M. and {Ng}, Alan and {Norris}, Ray P. and {Phillips}, Chris and {Reynolds}, John E. and {Roxby}, Daniel N. and {Schinckel}, Antony E.~T. and {Shields}, Matt and {Tremblay}, Chenoa and {Tzioumis}, A. and {Voronkov}, M.~A. and {Westmeier}, Tobias},
        title = "{The Rapid ASKAP Continuum Survey I: Design and first results}",
      journal = {\pasa},
         year = 2020,
        month = nov,
       volume = {37},
          eid = {e048},
        pages = {e048},
          doi = {10.1017/pasa.2020.41},
archivePrefix = {arXiv},
       eprint = {2012.00747},
 primaryClass = {astro-ph.IM},
       adsurl = {https://ui.adsabs.harvard.edu/abs/2020PASA...37...48M}
}

@ARTICLE{Brunetti2014,
   author = {{Brunetti}, G. and {Jones}, T.~W.},
    title = "{Cosmic Rays in Galaxy Clusters and Their Nonthermal Emission}",
  journal = {International Journal of Modern Physics D},
archivePrefix = "arXiv",
   eprint = {1401.7519},
     year = 2014,
    month = mar,
   volume = 23,
      eid = {1430007-98},
    pages = {1430007-98},
      doi = {10.1142/S0218271814300079},
   adsurl = {http://adsabs.harvard.edu/abs/2014IJMPD..2330007B}
}

@ARTICLE{Rajpurohit2023,
       author = {{Rajpurohit}, K. and {Osinga}, E. and {Brienza}, M. and {Botteon}, A. and {Brunetti}, G. and {Forman}, W.~R. and {Riseley}, C.~J. and {Vazza}, F. and {Bonafede}, A. and {van Weeren}, R.~J. and {Br{\"u}ggen}, M. and {Rajpurohit}, S. and {Drabent}, A. and {Dallacasa}, D. and {Rossetti}, M. and {Rajpurohit}, A.~S. and {Hoeft}, M. and {Bonnassieux}, E. and {Cassano}, R. and {Miley}, G.~K.},
        title = "{Deep low-frequency radio observations of Abell 2256. II. The ultra-steep spectrum radio halo}",
      journal = {\aap},
         year = 2023,
        month = jan,
       volume = {669},
          eid = {A1},
        pages = {A1},
          doi = {10.1051/0004-6361/202244925},
archivePrefix = {arXiv},
       eprint = {2209.03288},
 primaryClass = {astro-ph.CO},
       adsurl = {https://ui.adsabs.harvard.edu/abs/2023A&A...669A...1R}
}

@ARTICLE{2013JCAP...07..022H,
       author = {{Harrison}, Ian and {Hotchkiss}, Shaun},
        title = "{A consistent approach to falsifying {\ensuremath{\Lambda}}CDM with rare galaxy clusters}",
      journal = {\jcap},
         year = 2013,
        month = jul,
       volume = {2013},
       number = {7},
          eid = {022},
        pages = {022},
          doi = {10.1088/1475-7516/2013/07/022},
archivePrefix = {arXiv},
       eprint = {1210.4369},
 primaryClass = {astro-ph.CO},
       adsurl = {https://ui.adsabs.harvard.edu/abs/2013JCAP...07..022H}
}

@ARTICLE{2011PhRvD..83b3015M,
       author = {{Mortonson}, Michael J. and {Hu}, Wayne and {Huterer}, Dragan},
        title = "{Simultaneous falsification of {\ensuremath{\Lambda}}CDM and quintessence with massive, distant clusters}",
      journal = {\prd},
         year = 2011,
        month = jan,
       volume = {83},
       number = {2},
          eid = {023015},
        pages = {023015},
          doi = {10.1103/PhysRevD.83.023015},
archivePrefix = {arXiv},
       eprint = {1011.0004},
 primaryClass = {astro-ph.CO},
       adsurl = {https://ui.adsabs.harvard.edu/abs/2011PhRvD..83b3015M}
}

@ARTICLE{Cuciti2021,
       author = {{Cuciti}, V. and {Cassano}, R. and {Brunetti}, G. and {Dallacasa}, D. and {de Gasperin}, F. and {Ettori}, S. and {Giacintucci}, S. and {Kale}, R. and {Pratt}, G.~W. and {van Weeren}, R.~J. and {Venturi}, T.},
        title = "{Radio halos in a mass-selected sample of 75 galaxy clusters. II. Statistical analysis}",
      journal = {\aap},
         year = 2021,
        month = mar,
       volume = {647},
          eid = {A51},
        pages = {A51},
          doi = {10.1051/0004-6361/202039208},
archivePrefix = {arXiv},
       eprint = {2101.01641},
 primaryClass = {astro-ph.CO},
       adsurl = {https://ui.adsabs.harvard.edu/abs/2021A&A...647A..51C}
}

@ARTICLE{Bonafede2022,
      author = {{Bonafede}, A. and {Brunetti}, G. and {Rudnick}, L. and {Vazza}, F. and {Bourdin}, H. and {Giovannini}, G. and {Shimwell}, T.~W. and {Zhang}, X. and {Mazzotta}, P. and {Simionescu}, A. and {Biava}, N. and {Bonnassieux}, E. and {Brienza}, M. and {Br{\"u}ggen}, M. and {Rajpurohit}, K. and {Riseley}, C.~J. and {Stuardi}, C. and {Feretti}, L. and {Tasse}, C. and {Botteon}, A. and {Carretti}, E. and {Cassano}, R. and {Cuciti}, V. and {Gasperin}, F. de and {Gastaldello}, F. and {Rossetti}, M. and {Rottgering}, H.~J.~A. and {Venturi}, T. and {Weeren}, R.~J. van},
        title = "{The Coma Cluster at LOFAR Frequencies. II. The Halo, Relic, and a New Accretion Relic}",
      journal = {\apj},
         year = 2022,
        month = jul,
       volume = {933},
       number = {2},
          eid = {218},
        pages = {218},
          doi = {10.3847/1538-4357/ac721d},
archivePrefix = {arXiv},
       eprint = {2203.01958},
 primaryClass = {astro-ph.HE},
       adsurl = {https://ui.adsabs.harvard.edu/abs/2022ApJ...933..218B}
}

@ARTICLE{2005Natur.435..629S,
       author = {{Springel}, Volker and {White}, Simon D.~M. and {Jenkins}, Adrian and {Frenk}, Carlos S. and {Yoshida}, Naoki and {Gao}, Liang and {Navarro}, Julio and {Thacker}, Robert and {Croton}, Darren and {Helly}, John and {Peacock}, John A. and {Cole}, Shaun and {Thomas}, Peter and {Couchman}, Hugh and {Evrard}, August and {Colberg}, J{\"o}rg and {Pearce}, Frazer},
        title = "{Simulations of the formation, evolution and clustering of galaxies and quasars}",
      journal = {\nat},
         year = 2005,
        month = jun,
       volume = {435},
       number = {7042},
        pages = {629-636},
          doi = {10.1038/nature03597},
archivePrefix = {arXiv},
       eprint = {astro-ph/0504097},
 primaryClass = {astro-ph},
       adsurl = {https://ui.adsabs.harvard.edu/abs/2005Natur.435..629S}
}

@ARTICLE{2008ApJ...688..709T,
       author = {{Tinker}, Jeremy and {Kravtsov}, Andrey V. and {Klypin}, Anatoly and {Abazajian}, Kevork and {Warren}, Michael and {Yepes}, Gustavo and {Gottl{\"o}ber}, Stefan and {Holz}, Daniel E.},
        title = "{Toward a Halo Mass Function for Precision Cosmology: The Limits of Universality}",
      journal = {\apj},
         year = 2008,
        month = dec,
       volume = {688},
       number = {2},
        pages = {709-728},
          doi = {10.1086/591439},
archivePrefix = {arXiv},
       eprint = {0803.2706},
 primaryClass = {astro-ph},
       adsurl = {https://ui.adsabs.harvard.edu/abs/2008ApJ...688..709T}
}

@ARTICLE{Pearce2017,
       author = {{Pearce}, C.~J.~J. and {van Weeren}, R.~J. and {Andrade-Santos}, F. and {Jones}, C. and {Forman}, W.~R. and {Br{\"u}ggen}, M. and {Bulbul}, E. and {Clarke}, T.~E. and {Kraft}, R.~P. and {Medezinski}, E. and {Mroczkowski}, T. and {Nonino}, M. and {Nulsen}, P.~E.~J. and {Randall}, S.~W. and {Umetsu}, K.},
        title = "{VLA Radio Observations of the HST Frontier Fields Cluster Abell 2744: The Discovery of New Radio Relics}",
      journal = {\apj},
         year = 2017,
        month = aug,
       volume = {845},
       number = {1},
          eid = {81},
        pages = {81},
          doi = {10.3847/1538-4357/aa7e2f},
archivePrefix = {arXiv},
       eprint = {1708.03367},
 primaryClass = {astro-ph.CO},
       adsurl = {https://ui.adsabs.harvard.edu/abs/2017ApJ...845...81P}
}

@ARTICLE{Rajpurohit2025,
       author = {{Rajpurohit}, K. and {Botteon}, A. and {O'Sullivan}, E. and {Forman}, W. and {Balboni}, M. and {Bruno}, L. and {van Weeren}, R.~J. and {Hoeft}, M. and {Brunetti}, G. and {Jones}, C. and {Rajpurohit}, A.~S. and {Sikhosana}, S.~P.},
        title = "{Radial Profiles of Radio Halos in Massive Galaxy Clusters: Diffuse Giants Over 2 Mpc}",
      journal = {\apj},
         year = 2025,
        month = oct,
       volume = {992},
       number = {1},
          eid = {88},
        pages = {88},
          doi = {10.3847/1538-4357/adfd59},
archivePrefix = {arXiv},
       eprint = {2505.05415},
 primaryClass = {astro-ph.CO},
       adsurl = {https://ui.adsabs.harvard.edu/abs/2025ApJ...992...88R}
}

@ARTICLE{Rajpurohit2021b,
       author = {{Rajpurohit}, K. and {Brunetti}, G. and {Bonafede}, A. and {van Weeren}, R.~J. and {Botteon}, A. and {Vazza}, F. and {Hoeft}, M. and {Riseley}, C.~J. and {Bonnassieux}, E. and {Brienza}, M. and {Forman}, W.~R. and {R{\"o}ttgering}, H.~J.~A. and {Rajpurohit}, A.~S. and {Locatelli}, N. and {Shimwell}, T.~W. and {Cassano}, R. and {Di Gennaro}, G. and {Br{\"u}ggen}, M. and {Wittor}, D. and {Drabent}, A. and {Ignesti}, A.},
        title = "{Physical insights from the spectrum of the radio halo in MACS J0717.5+3745}",
      journal = {\aap},
         year = 2021,
        month = feb,
       volume = {646},
          eid = {A135},
        pages = {A135},
          doi = {10.1051/0004-6361/202039591},
       adsurl = {https://ui.adsabs.harvard.edu/abs/2021A&A...646A.135R}
}

@ARTICLE{2021A&A...648A...1T,
       author = {{Tasse}, C. and {Shimwell}, T. and {Hardcastle}, M.~J. and {O'Sullivan}, S.~P. and {van Weeren}, R. and {Best}, P.~N. and {Bester}, L. and {Hugo}, B. and {Smirnov}, O. and {Sabater}, J. and {Calistro-Rivera}, G. and {de Gasperin}, F. and {Morabito}, L.~K. and {R{\"o}ttgering}, H. and {Williams}, W.~L. and {Bonato}, M. and {Bondi}, M. and {Botteon}, A. and {Br{\"u}ggen}, M. and {Brunetti}, G. and {Chy{\.z}y}, K.~T. and {Garrett}, M.~A. and {G{\"u}rkan}, G. and {Jarvis}, M.~J. and {Kondapally}, R. and {Mandal}, S. and {Prandoni}, I. and {Repetti}, A. and {Retana-Montenegro}, E. and {Schwarz}, D.~J. and {Shulevski}, A. and {Wiaux}, Y.},
        title = "{The LOFAR Two-meter Sky Survey: Deep Fields Data Release 1. I. Direction-dependent calibration and imaging}",
      journal = {\aap},
         year = 2021,
        month = apr,
       volume = {648},
          eid = {A1},
        pages = {A1},
          doi = {10.1051/0004-6361/202038804},
archivePrefix = {arXiv},
       eprint = {2011.08328},
 primaryClass = {astro-ph.IM},
       adsurl = {https://ui.adsabs.harvard.edu/abs/2021A&A...648A...1T}
}

@ARTICLE{2022A&A...660A..78B,
       author = {{Botteon}, A. and {Shimwell}, T.~W. and {Cassano}, R. and {Cuciti}, V. and {Zhang}, X. and {Bruno}, L. and {Camillini}, L. and {Natale}, R. and {Jones}, A. and {Gastaldello}, F. and {Simionescu}, A. and {Rossetti}, M. and {Akamatsu}, H. and {van Weeren}, R.~J. and {Brunetti}, G. and {Br{\"u}ggen}, M. and {Groeneveld}, C. and {Hoang}, D.~N. and {Hardcastle}, M.~J. and {Ignesti}, A. and {Di Gennaro}, G. and {Bonafede}, A. and {Drabent}, A. and {R{\"o}ttgering}, H.~J.~A. and {Hoeft}, M. and {de Gasperin}, F.},
        title = "{The Planck clusters in the LOFAR sky. I. LoTSS-DR2: New detections and sample overview}",
      journal = {\aap},
         year = 2022,
        month = apr,
       volume = {660},
          eid = {A78},
        pages = {A78},
          doi = {10.1051/0004-6361/202143020},
archivePrefix = {arXiv},
       eprint = {2202.11720},
 primaryClass = {astro-ph.CO},
       adsurl = {https://ui.adsabs.harvard.edu/abs/2022A&A...660A..78B}
}

@ARTICLE{2026MNRAS.546ag054V,
       author = {{van Weeren}, R.~J. and {Osinga}, E. and {Brunetti}, G. and {Riseley}, C.~J. and {Botteon}, A. and {Timmerman}, R. and {Bonafede}, A. and {Br{\"u}ggen}, M. and {Cassano}, R. and {Cuciti}, V. and {Dallacasa}, D. and {de Gasperin}, F. and {de Jong}, J.~M.~G.~H.~J. and {Gastaldello}, F. and {Knowles}, K. and {Zhang}, X.},
        title = "{MeerKAT observations of Abell 1775 and Abell 1795: the discovery of a hadronic radio halo?}",
      journal = {\mnras},
         year = 2026,
        month = feb,
       volume = {546},
       number = {2},
          eid = {stag054},
        pages = {stag054},
          doi = {10.1093/mnras/stag054},
archivePrefix = {arXiv},
       eprint = {2601.16288},
 primaryClass = {astro-ph.CO},
       adsurl = {https://ui.adsabs.harvard.edu/abs/2026MNRAS.546ag054V}
}

@ARTICLE{2021A&A...651A.115V,
       author = {{van Weeren}, R.~J. and {Shimwell}, T.~W. and {Botteon}, A. and {Brunetti}, G. and {Br{\"u}ggen}, M. and {Boxelaar}, J.~M. and {Cassano}, R. and {Di Gennaro}, G. and {Andrade-Santos}, F. and {Bonnassieux}, E. and {Bonafede}, A. and {Cuciti}, V. and {Dallacasa}, D. and {de Gasperin}, F. and {Gastaldello}, F. and {Hardcastle}, M.~J. and {Hoeft}, M. and {Kraft}, R.~P. and {Mandal}, S. and {Rossetti}, M. and {R{\"o}ttgering}, H.~J.~A. and {Tasse}, C. and {Wilber}, A.~G.},
        title = "{LOFAR observations of galaxy clusters in HETDEX. Extraction and self-calibration of individual LOFAR targets}",
      journal = {\aap},
         year = 2021,
        month = jul,
       volume = {651},
          eid = {A115},
        pages = {A115},
          doi = {10.1051/0004-6361/202039826},
archivePrefix = {arXiv},
       eprint = {2011.02387},
 primaryClass = {astro-ph.CO},
       adsurl = {https://ui.adsabs.harvard.edu/abs/2021A&A...651A.115V}
}

@ARTICLE{vanWeeren2019,
       author = {{van Weeren}, R.~J. and {de Gasperin}, F. and {Akamatsu}, H. and
         {Br{\"u}ggen}, M. and {Feretti}, L. and {Kang}, H. and {Stroe}, A. and
         {Zandanel}, F.},
        title = "{Diffuse Radio Emission from Galaxy Clusters}",
      journal = {\ssr},
         year = "2019",
        month = "Feb",
       volume = {215},
       number = {1},
          eid = {16},
        pages = {16},
          doi = {10.1007/s11214-019-0584-z},
archivePrefix = {arXiv},
       eprint = {1901.04496},
 primaryClass = {astro-ph.HE},
       adsurl = {https://ui.adsabs.harvard.edu/abs/2019SSRv..215...16V}
}

@INPROCEEDINGS{Jonas2916,
       author = {{Jonas}, J. and {MeerKAT Team}},
        title = "{The MeerKAT Radio Telescope}",
    booktitle = {MeerKAT Science: On the Pathway to the SKA},
         year = 2016,
        month = jan,
          eid = {1},
        pages = {1},
          doi = {10.22323/1.277.0001},
       adsurl = {https://ui.adsabs.harvard.edu/abs/2016mks..confE...1J}
}

@ARTICLE{Haarlem2013,
   author = {{van Haarlem}, M.~P. and {Wise}, M.~W. and {Gunst}, A.~W. and 
	{Heald}, G. and {McKean}, J.~P. and {Hessels}, J.~W.~T. and 
	{de Bruyn}, A.~G. and {Nijboer}, R. and {Swinbank}, J. and {Fallows}, R. and 
	{Brentjens}, M. and {Nelles}, A. and {Beck}, R. and {Falcke}, H. and 
	{Fender}, R. and {H{\"o}randel}, J. and {Koopmans}, L.~V.~E. and 
	{Mann}, G. and {Miley}, G. and {R{\"o}ttgering}, H. and {Stappers}, B.~W. and 
	{Wijers}, R.~A.~M.~J. and {Zaroubi}, S. and {van den Akker}, M. and 
	{Alexov}, A. and {Anderson}, J. and {Anderson}, K. and {van Ardenne}, A. and 
	{Arts}, M. and {Asgekar}, A. and {Avruch}, I.~M. and {Batejat}, F. and 
	{B{\"a}hren}, L. and {Bell}, M.~E. and {Bell}, M.~R. and {van Bemmel}, I. and 
	{Bennema}, P. and {Bentum}, M.~J. and {Bernardi}, G. and {Best}, P. and 
	{B{\^i}rzan}, L. and {Bonafede}, A. and {Boonstra}, A.-J. and 
	{Braun}, R. and {Bregman}, J. and {Breitling}, F. and {van de Brink}, R.~H. and 
	{Broderick}, J. and {Broekema}, P.~C. and {Brouw}, W.~N. and 
	{Br{\"u}ggen}, M. and {Butcher}, H.~R. and {van Cappellen}, W. and 
	{Ciardi}, B. and {Coenen}, T. and {Conway}, J. and {Coolen}, A. and 
	{Corstanje}, A. and {Damstra}, S. and {Davies}, O. and {Deller}, A.~T. and 
	{Dettmar}, R.-J. and {van Diepen}, G. and {Dijkstra}, K. and 
	{Donker}, P. and {Doorduin}, A. and {Dromer}, J. and {Drost}, M. and 
	{van Duin}, A. and {Eisl{\"o}ffel}, J. and {van Enst}, J. and 
	{Ferrari}, C. and {Frieswijk}, W. and {Gankema}, H. and {Garrett}, M.~A. and 
	{de Gasperin}, F. and {Gerbers}, M. and {de Geus}, E. and {Grie{\ss}meier}, J.-M. and {Grit}, T. and {Gruppen}, P. and {Hamaker}, J.~P. and {Hassall}, T. and 
	{Hoeft}, M. and {Holties}, H.~A. and {Horneffer}, A. and {van der Horst}, A. and 
	{van Houwelingen}, A. and {Huijgen}, A. and {Iacobelli}, M. and 
	{Intema}, H. and {Jackson}, N. and {Jelic}, V. and {de Jong}, A. and 
	{Juette}, E. and {Kant}, D. and {Karastergiou}, A. and {Koers}, A. and 
	{Kollen}, H. and {Kondratiev}, V.~I. and {Kooistra}, E. and 
	{Koopman}, Y. and {Koster}, A. and {Kuniyoshi}, M. and {Kramer}, M. and 
	{Kuper}, G. and {Lambropoulos}, P. and {Law}, C. and {van Leeuwen}, J. and 
	{Lemaitre}, J. and {Loose}, M. and {Maat}, P. and {Macario}, G. and 
	{Markoff}, S. and {Masters}, J. and {McFadden}, R.~A. and {McKay-Bukowski}, D. and 
	{Meijering}, H. and {Meulman}, H. and {Mevius}, M. and {Middelberg}, E. and 
	{Millenaar}, R. and {Miller-Jones}, J.~C.~A. and {Mohan}, R.~N. and 
	{Mol}, J.~D. and {Morawietz}, J. and {Morganti}, R. and {Mulcahy}, D.~D. and 
	{Mulder}, E. and {Munk}, H. and {Nieuwenhuis}, L. and {van Nieuwpoort}, R. and 
	{Noordam}, J.~E. and {Norden}, M. and {Noutsos}, A. and {Offringa}, A.~R. and 
	{Olofsson}, H. and {Omar}, A. and {Orr{\'u}}, E. and {Overeem}, R. and 
	{Paas}, H. and {Pandey-Pommier}, M. and {Pandey}, V.~N. and 
	{Pizzo}, R. and {Polatidis}, A. and {Rafferty}, D. and {Rawlings}, S. and 
	{Reich}, W. and {de Reijer}, J.-P. and {Reitsma}, J. and {Renting}, G.~A. and 
	{Riemers}, P. and {Rol}, E. and {Romein}, J.~W. and {Roosjen}, J. and 
	{Ruiter}, M. and {Scaife}, A. and {van der Schaaf}, K. and {Scheers}, B. and 
	{Schellart}, P. and {Schoenmakers}, A. and {Schoonderbeek}, G. and 
	{Serylak}, M. and {Shulevski}, A. and {Sluman}, J. and {Smirnov}, O. and 
	{Sobey}, C. and {Spreeuw}, H. and {Steinmetz}, M. and {Sterks}, C.~G.~M. and 
	{Stiepel}, H.-J. and {Stuurwold}, K. and {Tagger}, M. and {Tang}, Y. and 
	{Tasse}, C. and {Thomas}, I. and {Thoudam}, S. and {Toribio}, M.~C. and 
	{van der Tol}, B. and {Usov}, O. and {van Veelen}, M. and {van der Veen}, A.-J. and 
	{ter Veen}, S. and {Verbiest}, J.~P.~W. and {Vermeulen}, R. and 
	{Vermaas}, N. and {Vocks}, C. and {Vogt}, C. and {de Vos}, M. and 
	{van der Wal}, E. and {van Weeren}, R. and {Weggemans}, H. and 
	{Weltevrede}, P. and {White}, S. and {Wijnholds}, S.~J. and 
	{Wilhelmsson}, T. and {Wucknitz}, O. and {Yatawatta}, S. and 
	{Zarka}, P. and {Zensus}, A. and {van Zwieten}, J.},
    title = "{LOFAR: The LOw-Frequency ARray}",
  journal = {\aap},
archivePrefix = "arXiv",
   eprint = {1305.3550},
 primaryClass = "astro-ph.IM",
     year = 2013,
    month = aug,
   volume = 556,
      eid = {A2},
    pages = {A2},
      doi = {10.1051/0004-6361/201220873},
   adsurl = {http://adsabs.harvard.edu/abs/2013A%26A...556A...2V}
}

@ARTICLE{Shimwell2022,
       author = {{Shimwell}, T.~W. and {Hardcastle}, M.~J. and {Tasse}, C. and {Best}, P.~N. and {R{\"o}ttgering}, H.~J.~A. and {Williams}, W.~L. and {Botteon}, A. and {Drabent}, A. and {Mechev}, A. and {Shulevski}, A. and {van Weeren}, R.~J. and {Bester}, L. and {Br{\"u}ggen}, M. and {Brunetti}, G. and {Callingham}, J.~R. and {Chy{\.z}y}, K.~T. and {Conway}, J.~E. and {Dijkema}, T.~J. and {Duncan}, K. and {de Gasperin}, F. and {Hale}, C.~L. and {Haverkorn}, M. and {Hugo}, B. and {Jackson}, N. and {Mevius}, M. and {Miley}, G.~K. and {Morabito}, L.~K. and {Morganti}, R. and {Offringa}, A. and {Oonk}, J.~B.~R. and {Rafferty}, D. and {Sabater}, J. and {Smith}, D.~J.~B. and {Schwarz}, D.~J. and {Smirnov}, O. and {O'Sullivan}, S.~P. and {Vedantham}, H. and {White}, G.~J. and {Albert}, J.~G. and {Alegre}, L. and {Asabere}, B. and {Bacon}, D.~J. and {Bonafede}, A. and {Bonnassieux}, E. and {Brienza}, M. and {Bilicki}, M. and {Bonato}, M. and {Calistro Rivera}, G. and {Cassano}, R. and {Cochrane}, R. and {Croston}, J.~H. and {Cuciti}, V. and {Dallacasa}, D. and {Danezi}, A. and {Dettmar}, R.~J. and {Di Gennaro}, G. and {Edler}, H.~W. and {En{\ss}lin}, T.~A. and {Emig}, K.~L. and {Franzen}, T.~M.~O. and {Garc{\'\i}a-Vergara}, C. and {Grange}, Y.~G. and {G{\"u}rkan}, G. and {Hajduk}, M. and {Heald}, G. and {Heesen}, V. and {Hoang}, D.~N. and {Hoeft}, M. and {Horellou}, C. and {Iacobelli}, M. and {Jamrozy}, M. and {Jeli{\'c}}, V. and {Kondapally}, R. and {Kukreti}, P. and {Kunert-Bajraszewska}, M. and {Magliocchetti}, M. and {Mahatma}, V. and {Ma{\l}ek}, K. and {Mandal}, S. and {Massaro}, F. and {Meyer-Zhao}, Z. and {Mingo}, B. and {Mostert}, R.~I.~J. and {Nair}, D.~G. and {Nakoneczny}, S.~J. and {Nikiel-Wroczy{\'n}ski}, B. and {Orr{\'u}}, E. and {Pajdosz-{\'S}mierciak}, U. and {Pasini}, T. and {Prandoni}, I. and {van Piggelen}, H.~E. and {Rajpurohit}, K. and {Retana-Montenegro}, E. and {Riseley}, C.~J. and {Rowlinson}, A. and {Saxena}, A. and {Schrijvers}, C. and {Sweijen}, F. and {Siewert}, T.~M. and {Timmerman}, R. and {Vaccari}, M. and {Vink}, J. and {West}, J.~L. and {Wo{\l}owska}, A. and {Zhang}, X. and {Zheng}, J.},
        title = "{The LOFAR Two-metre Sky Survey. V. Second data release}",
      journal = {\aap},
         year = 2022,
        month = mar,
       volume = {659},
          eid = {A1},
        pages = {A1},
          doi = {10.1051/0004-6361/202142484},
archivePrefix = {arXiv},
       eprint = {2202.11733},
 primaryClass = {astro-ph.GA},
       adsurl = {https://ui.adsabs.harvard.edu/abs/2022A&A...659A...1S}
}

@ARTICLE{Rajpurohit2021c,
       author = {{Rajpurohit}, K. and {Vazza}, F. and {van Weeren}, R.~J. and {Hoeft}, M. and {Brienza}, M. and {Bonnassieux}, E. and {Riseley}, C.~J. and {Brunetti}, G. and {Bonafede}, A. and {Br{\"u}ggen}, M. and {Formann}, W.~R. and {Rajpurohit}, A.~S. and {R{\"o}ttgering}, H.~J.~A. and {Drabent}, A. and {Dom{\'\i}nguez-Fern{\'a}ndez}, P. and {Wittor}, D. and {Andrade-Santos}, F.},
        title = "{Dissecting nonthermal emission in the complex multiple-merger galaxy cluster Abell 2744: Radio and X-ray analysis}",
      journal = {\aap},
         year = 2021,
        month = oct,
       volume = {654},
          eid = {A41},
        pages = {A41},
          doi = {10.1051/0004-6361/202141060},
archivePrefix = {arXiv},
       eprint = {2104.05690},
 primaryClass = {astro-ph.CO},
       adsurl = {https://ui.adsabs.harvard.edu/abs/2021A&A...654A..41R}
}

@ARTICLE{Offringa2010,
       author = {{Offringa}, A.~R. and {de Bruyn}, A.~G. and {Biehl}, M. and
         {Zaroubi}, S. and {Bernardi}, G. and {Pandey}, V.~N.},
        title = "{Post-correlation radio frequency interference classification methods}",
      journal = {\mnras},
         year = "2010",
        month = "Jun",
       volume = {405},
       number = {1},
        pages = {155-167},
          doi = {10.1111/j.1365-2966.2010.16471.x},
archivePrefix = {arXiv},
       eprint = {1002.1957},
 primaryClass = {astro-ph.IM},
       adsurl = {https://ui.adsabs.harvard.edu/abs/2010MNRAS.405..155O}
}

@ARTICLE{Offringa2014,
       author = {{Offringa}, A.~R. and {McKinley}, B. and {Hurley-Walker}, N. and
         {Briggs}, F.~H. and {Wayth}, R.~B. and {Kaplan}, D.~L. and
         {Bell}, M.~E. and {Feng}, L. and {Neben}, A.~R. and {Hughes}, J.~D. and
         {Rhee}, J. and {Murphy}, T. and {Bhat}, N.~D.~R. and {Bernardi}, G. and
         {Bowman}, J.~D. and {Cappallo}, R.~J. and {Corey}, B.~E. and {Deshpand
        e}, A.~A. and {Emrich}, D. and {Ewall-Wice}, A. and {Gaensler}, B.~M. and
         {Goeke}, R. and {Greenhill}, L.~J. and {Hazelton}, B.~J. and
         {Hindson}, L. and {Johnston-Hollitt}, M. and {Jacobs}, D.~C. and
         {Kasper}, J.~C. and {Kratzenberg}, E. and {Lenc}, E. and
         {Lonsdale}, C.~J. and {Lynch}, M.~J. and {McWhirter}, S.~R. and
         {Mitchell}, D.~A. and {Morales}, M.~F. and {Morgan}, E. and
         {Kudryavtseva}, N. and {Oberoi}, D. and {Ord}, S.~M. and {Pindor}, B. and
         {Procopio}, P. and {Prabu}, T. and {Riding}, J. and {Roshi}, D.~A. and
         {Shankar}, N. Udaya and {Srivani}, K.~S. and {Subrahmanyan}, R. and
         {Tingay}, S.~J. and {Waterson}, M. and {Webster}, R.~L. and
         {Whitney}, A.~R. and {Williams}, A. and {Williams}, C.~L.},
        title = "{WSCLEAN: an implementation of a fast, generic wide-field imager for radio astronomy}",
      journal = {\mnras},
         year = 2014,
        month = oct,
       volume = {444},
       number = {1},
        pages = {606-619},
          doi = {10.1093/mnras/stu1368},
archivePrefix = {arXiv},
       eprint = {1407.1943},
 primaryClass = {astro-ph.IM},
       adsurl = {https://ui.adsabs.harvard.edu/abs/2014MNRAS.444..606O}
}

@ARTICLE{Kenyon2018,
       author = {{Kenyon}, J.~S. and {Smirnov}, O.~M. and {Grobler}, T.~L. and {Perkins}, S.~J.},
        title = "{CUBICAL - fast radio interferometric calibration suite exploiting complex optimization}",
      journal = {\mnras},
         year = 2018,
        month = aug,
       volume = {478},
       number = {2},
        pages = {2399-2415},
          doi = {10.1093/mnras/sty1221},
archivePrefix = {arXiv},
       eprint = {1805.03410},
 primaryClass = {astro-ph.IM},
       adsurl = {https://ui.adsabs.harvard.edu/abs/2018MNRAS.478.2399K}
}

@ARTICLE{Scaife2012,
       author = {{Scaife}, Anna M.~M. and {Heald}, George H.},
        title = "{A broad-band flux scale for low-frequency radio telescopes}",
      journal = {\mnras},
         year = 2012,
        month = jun,
       volume = {423},
       number = {1},
        pages = {L30-L34},
          doi = {10.1111/j.1745-3933.2012.01251.x},
archivePrefix = {arXiv},
       eprint = {1203.0977},
 primaryClass = {astro-ph.IM},
       adsurl = {https://ui.adsabs.harvard.edu/abs/2012MNRAS.423L..30S}
}

@ARTICLE{Intema2009,
       author = {{Intema}, H.~T. and {van der Tol}, S. and {Cotton}, W.~D. and {Cohen}, A.~S. and {van Bemmel}, I.~M. and {R{\"o}ttgering}, H.~J.~A.},
        title = "{Ionospheric calibration of low frequency radio interferometric observations using the peeling scheme. I. Method description and first results}",
      journal = {\aap},
         year = 2009,
        month = jul,
       volume = {501},
       number = {3},
        pages = {1185-1205},
          doi = {10.1051/0004-6361/200811094},
archivePrefix = {arXiv},
       eprint = {0904.3975},
 primaryClass = {astro-ph.IM},
       adsurl = {https://ui.adsabs.harvard.edu/abs/2009A&A...501.1185I}
}

@MISC{caracal2020,
       author = {{J{\'o}zsa}, Gyula I.~G. and {White}, Sarah V. and {Thorat}, Kshitij and {Smirnov}, Oleg M. and {Serra}, Paolo and {Ramatsoku}, Mpati and {Ramaila}, Athanaseus J.~T. and {Perkins}, Simon J. and {Moln{\'a}r}, D{\'a}niel Cs. and {Makhathini}, Sphesihle and {Maccagni}, Filippo M. and {Kleiner}, Dane and {Kamphuis}, Peter and {Hugo}, Benjamin V. and {de Blok}, W.~J.~G. and {Andati}, Lexy A.~L.},
        title = "{CARACal: Containerized Automated Radio Astronomy Calibration pipeline}",
 howpublished = {Astrophysics Source Code Library, record ascl:2006.014},
         year = 2020,
        month = jun,
          eid = {ascl:2006.014},
        pages = {ascl:2006.014},
archivePrefix = {ascl},
       eprint = {2006.014},
       adsurl = {https://ui.adsabs.harvard.edu/abs/2020ascl.soft06014J}
}

@ARTICLE{Lyskova2025,
       author = {{Lyskova}, N. and {Churazov}, E. and {Khabibullin}, I. and {Bikmaev}, I.~F. and {Burenin}, R.~A. and {Forman}, W.~R. and {Khamitov}, I.~M. and {Rajpurohit}, K. and {Sunyaev}, R. and {Jones}, C. and {Kraft}, R. and {Zaznobin}, I. and {Gorbachev}, M.~A. and {Suslikov}, M.~V. and {Gumerov}, R.~I. and {Sakhibullin}, N.~A.},
        title = "{The massive galaxy cluster CL0238.3+2005 (the Peanut cluster) at z = 0.42: A merger just after pericenter passage?}",
      journal = {\aap},
         year = 2025,
        month = jan,
       volume = {693},
          eid = {A55},
        pages = {A55},
          doi = {10.1051/0004-6361/202452212},
archivePrefix = {arXiv},
       eprint = {2409.07856},
 primaryClass = {astro-ph.CO},
       adsurl = {https://ui.adsabs.harvard.edu/abs/2025A&A...693A..55L}
}

@ARTICLE{2020ApJ...891....1G,
       author = {{Giacintucci}, S. and {Markevitch}, M. and {Johnston-Hollitt}, M. and {Wik}, D.~R. and {Wang}, Q.~H.~S. and {Clarke}, T.~E.},
        title = "{Discovery of a Giant Radio Fossil in the Ophiuchus Galaxy Cluster}",
      journal = {\apj},
         year = 2020,
        month = mar,
       volume = {891},
       number = {1},
          eid = {1},
        pages = {1},
          doi = {10.3847/1538-4357/ab6a9d},
archivePrefix = {arXiv},
       eprint = {2002.01291},
 primaryClass = {astro-ph.GA},
       adsurl = {https://ui.adsabs.harvard.edu/abs/2020ApJ...891....1G}
}

@ARTICLE{2026arXiv260427161H,
       author = {{Heinrich}, Annie and {Zhuravleva}, Irina and {Churazov}, Eugene and {Zhang}, Congyao and {Rogantini}, Daniele and {McCall}, Hannah and {van Weeren}, Reinout J. and {Forman}, William R.},
        title = "{The major cluster merger in Abell 2034 as seen by XRISM: Strong turbulence and spectral anomalies?}",
      journal = {arXiv e-prints},
         year = 2026,
        month = apr,
          eid = {arXiv:2604.27161},
        pages = {arXiv:2604.27161},
          doi = {10.48550/arXiv.2604.27161},
archivePrefix = {arXiv},
       eprint = {2604.27161},
 primaryClass = {astro-ph.HE},
       adsurl = {https://ui.adsabs.harvard.edu/abs/2026arXiv260427161H}
}

@ARTICLE{2024MNRAS.528.6470M,
       author = {{Murgia}, M. and {Govoni}, F. and {Vacca}, V. and {Loi}, F. and {Feretti}, L. and {Giovannini}, G. and {Melis}, A. and {Concu}, R. and {Carretti}, E. and {Poppi}, S. and {Valente}, G. and {Bernardi}, G. and {Bonafede}, A. and {Boschin}, W. and {Brienza}, M. and {Clarke}, T.~E. and {de Gasperin}, F. and {En{\ss}lin}, T.~A. and {Ferrari}, C. and {Gastaldello}, F. and {Girardi}, M. and {Gregorini}, L. and {Johnston-Hollitt}, M. and {Orr{\`u}}, E. and {Parma}, P. and {Perley}, R.~A. and {Taylor}, G.~B. and {Marchegiani}, P.},
        title = "{Sardinia Radio Telescope observations of the Coma cluster}",
      journal = {\mnras},
         year = 2024,
        month = mar,
       volume = {528},
       number = {4},
        pages = {6470-6495},
          doi = {10.1093/mnras/stae436},
archivePrefix = {arXiv},
       eprint = {2402.07306},
 primaryClass = {astro-ph.CO},
       adsurl = {https://ui.adsabs.harvard.edu/abs/2024MNRAS.528.6470M}
}

@ARTICLE{2024A&A...692A..12V,
       author = {{van Weeren}, R.~J. and {Timmerman}, R. and {Vaidya}, V. and {Gendron-Marsolais}, M.-L. and {Botteon}, A. and {Roberts}, I.~D. and {Hlavacek-Larrondo}, J. and {Bonafede}, A. and {Br{\"u}ggen}, M. and {Brunetti}, G. and {Cassano}, R. and {Cuciti}, V. and {Edge}, A.~C. and {Gastaldello}, F. and {Groeneveld}, C. and {Shimwell}, T.~W.},
        title = "{LOFAR high-band antenna observations of the Perseus cluster: The discovery of a giant radio halo}",
      journal = {\aap},
         year = 2024,
        month = dec,
       volume = {692},
          eid = {A12},
        pages = {A12},
          doi = {10.1051/0004-6361/202451618},
archivePrefix = {arXiv},
       eprint = {2410.02863},
 primaryClass = {astro-ph.CO},
       adsurl = {https://ui.adsabs.harvard.edu/abs/2024A&A...692A..12V}
}

@ARTICLE{1983ApJ...271..575B,
       author = {{Burns}, J.~O. and {Schwendeman}, E. and {White}, R.~A.},
        title = "{CD galaxy dynamics and an aged ridge (jet) in 3C 338.}",
      journal = {\apj},
         year = 1983,
        month = aug,
       volume = {271},
        pages = {575-585},
          doi = {10.1086/161224},
       adsurl = {https://ui.adsabs.harvard.edu/abs/1983ApJ...271..575B}
}

@ARTICLE{2013ApJ...775..117N,
       author = {{Nulsen}, Paul E.~J. and {Li}, Zhiyuan and {Forman}, William R. and {Kraft}, Ralph P. and {Lal}, Dharam V. and {Jones}, Christine and {Zhuravleva}, Irina and {Churazov}, Eugene and {Sanders}, Jeremy S. and {Fabian}, Andrew C. and {Johnson}, Ryan E. and {Murray}, Stephen S.},
        title = "{Deep Chandra Observations of A2199: The Interplay between Merger-induced Gas Motions and Nuclear Outbursts in a Cool Core Cluster}",
      journal = {\apj},
         year = 2013,
        month = oct,
       volume = {775},
       number = {2},
          eid = {117},
        pages = {117},
          doi = {10.1088/0004-637X/775/2/117},
archivePrefix = {arXiv},
       eprint = {1307.8072},
 primaryClass = {astro-ph.CO},
       adsurl = {https://ui.adsabs.harvard.edu/abs/2013ApJ...775..117N}
}

@ARTICLE{2016MNRAS.459..277S,
       author = {{Shimwell}, T.~W. and {Luckin}, J. and {Br{\"u}ggen}, M. and {Brunetti}, G. and {Intema}, H.~T. and {Owers}, M.~S. and {R{\"o}ttgering}, H.~J.~A. and {Stroe}, A. and {van Weeren}, R.~J. and {Williams}, W.~L. and {Cassano}, R. and {de Gasperin}, F. and {Heald}, G.~H. and {Hoang}, D.~N. and {Hardcastle}, M.~J. and {Sridhar}, S.~S. and {Sabater}, J. and {Best}, P.~N. and {Bonafede}, A. and {Chy{\.z}y}, K.~T. and {En{\ss}lin}, T.~A. and {Ferrari}, C. and {Haverkorn}, M. and {Hoeft}, M. and {Horellou}, C. and {McKean}, J.~P. and {Morabito}, L.~K. and {Orr{\`u}}, E. and {Pizzo}, R. and {Retana-Montenegro}, E. and {White}, G.~J.},
        title = "{A plethora of diffuse steep spectrum radio sources in Abell 2034 revealed by LOFAR}",
      journal = {\mnras},
         year = 2016,
        month = jun,
       volume = {459},
       number = {1},
        pages = {277-290},
          doi = {10.1093/mnras/stw661},
archivePrefix = {arXiv},
       eprint = {1603.06591},
 primaryClass = {astro-ph.HE},
       adsurl = {https://ui.adsabs.harvard.edu/abs/2016MNRAS.459..277S}
}

@ARTICLE{2002MNRAS.331.1011E,
       author = {{En{\ss}lin}, T.~A. and {Br{\"u}ggen}, M.},
        title = "{On the formation of cluster radio relics}",
      journal = {\mnras},
         year = 2002,
        month = apr,
       volume = {331},
       number = {4},
        pages = {1011-1019},
          doi = {10.1046/j.1365-8711.2002.05261.x},
archivePrefix = {arXiv},
       eprint = {astro-ph/0104233},
 primaryClass = {astro-ph},
       adsurl = {https://ui.adsabs.harvard.edu/abs/2002MNRAS.331.1011E}
}

@ARTICLE{2005ApJ...634L.141H,
       author = {{Heinz}, S. and {Churazov}, E.},
        title = "{Heating the Bubbly Gas of Galaxy Clusters with Weak Shocks and Sound Waves}",
      journal = {\apjl},
         year = 2005,
        month = dec,
       volume = {634},
       number = {2},
        pages = {L141-L144},
          doi = {10.1086/498301},
archivePrefix = {arXiv},
       eprint = {astro-ph/0507038},
 primaryClass = {astro-ph},
       adsurl = {https://ui.adsabs.harvard.edu/abs/2005ApJ...634L.141H}
}

@ARTICLE{1987JFM...181...41H,
       author = {{Haas}, J.-F. and {Sturtevant}, B.},
        title = "{Interaction of weak shock waves with cylindrical and spherical gas inhomogeneities}",
      journal = {Journal of Fluid Mechanics},
         year = 1987,
        month = aug,
       volume = {181},
        pages = {41-76},
          doi = {10.1017/S0022112087002003},
       adsurl = {https://ui.adsabs.harvard.edu/abs/1987JFM...181...41H}
}

@ARTICLE{2011AnRFM..43..117R,
       author = {{Ranjan}, Devesh and {Oakley}, Jason and {Bonazza}, Riccardo},
        title = "{Shock-Bubble Interactions}",
      journal = {Annual Review of Fluid Mechanics},
         year = 2011,
        month = jan,
       volume = {43},
       number = {1},
        pages = {117-140},
          doi = {10.1146/annurev-fluid-122109-160744},
       adsurl = {https://ui.adsabs.harvard.edu/abs/2011AnRFM..43..117R}
}

@ARTICLE{2026A&A...705A.184C,
       author = {{Churazov}, Eugene and {Rudnick}, Lawrence and {Khabibullin}, Ildar I. and {Brienza}, Marisa and {Schekochihin}, Alex and {Uzdensky}, Dmitri},
        title = "{Fast and ``lossless'' propagation of relativistic electrons along magnetized nonthermal filaments in galaxy clusters and the Galactic Center region}",
      journal = {\aap},
         year = 2026,
        month = jan,
       volume = {705},
          eid = {A184},
        pages = {A184},
          doi = {10.1051/0004-6361/202556619},
archivePrefix = {arXiv},
       eprint = {2507.20780},
 primaryClass = {astro-ph.HE},
       adsurl = {https://ui.adsabs.harvard.edu/abs/2026A&A...705A.184C}
}

@ARTICLE{2019SSRv..215...17M,
       author = {{Mroczkowski}, Tony and {Nagai}, Daisuke and {Basu}, Kaustuv and {Chluba}, Jens and {Sayers}, Jack and {Adam}, R{\'e}mi and {Churazov}, Eugene and {Crites}, Abigail and {Di Mascolo}, Luca and {Eckert}, Dominique and {Macias-Perez}, Juan and {Mayet}, Fr{\'e}d{\'e}ric and {Perotto}, Laurence and {Pointecouteau}, Etienne and {Romero}, Charles and {Ruppin}, Florian and {Scannapieco}, Evan and {ZuHone}, John},
        title = "{Astrophysics with the Spatially and Spectrally Resolved Sunyaev-Zeldovich Effects. A Millimetre/Submillimetre Probe of the Warm and Hot Universe}",
      journal = {\ssr},
         year = 2019,
        month = feb,
       volume = {215},
       number = {1},
          eid = {17},
        pages = {17},
          doi = {10.1007/s11214-019-0581-2},
archivePrefix = {arXiv},
       eprint = {1811.02310},
 primaryClass = {astro-ph.CO},
       adsurl = {https://ui.adsabs.harvard.edu/abs/2019SSRv..215...17M}
}

@ARTICLE{Intema2017,
       author = {{Intema}, H.~T. and {Jagannathan}, P. and {Mooley}, K.~P. and {Frail}, D.~A.},
        title = "{The GMRT 150 MHz all-sky radio survey. First alternative data release TGSS ADR1}",
      journal = {\aap},
         year = 2017,
        month = feb,
       volume = {598},
          eid = {A78},
        pages = {A78},
          doi = {10.1051/0004-6361/201628536},
archivePrefix = {arXiv},
       eprint = {1603.04368},
 primaryClass = {astro-ph.CO},
       adsurl = {https://ui.adsabs.harvard.edu/abs/2017A&A...598A..78I}
}

@ARTICLE{2004ApJ...606..819M,
       author = {{Markevitch}, M. and {Gonzalez}, A.~H. and {Clowe}, D. and {Vikhlinin}, A. and {Forman}, W. and {Jones}, C. and {Murray}, S. and {Tucker}, W.},
        title = "{Direct Constraints on the Dark Matter Self-Interaction Cross Section from the Merging Galaxy Cluster 1E 0657-56}",
      journal = {\apj},
         year = 2004,
        month = may,
       volume = {606},
       number = {2},
        pages = {819-824},
          doi = {10.1086/383178},
archivePrefix = {arXiv},
       eprint = {astro-ph/0309303},
 primaryClass = {astro-ph},
       adsurl = {https://ui.adsabs.harvard.edu/abs/2004ApJ...606..819M}
}

@ARTICLE{2016A&A...594A.121P,
       author = {{Paraficz}, D. and {Kneib}, J. -P. and {Richard}, J. and {Morandi}, A. and {Limousin}, M. and {Jullo}, E. and {Martinez}, J.},
        title = "{The Bullet cluster at its best: weighing stars, gas, and dark matter}",
      journal = {\aap},
         year = 2016,
        month = oct,
       volume = {594},
          eid = {A121},
        pages = {A121},
          doi = {10.1051/0004-6361/201527959},
archivePrefix = {arXiv},
       eprint = {1209.0384},
 primaryClass = {astro-ph.CO},
       adsurl = {https://ui.adsabs.harvard.edu/abs/2016A&A...594A.121P}
}

@ARTICLE{2006A&A...451..395C,
       author = {{Clowe}, D. and {Schneider}, P. and {Arag{\'o}n-Salamanca}, A. and {Bremer}, M. and {De Lucia}, G. and {Halliday}, C. and {Jablonka}, P. and {Milvang-Jensen}, B. and {Pell{\'o}}, R. and {Poggianti}, B. and {Rudnick}, G. and {Saglia}, R. and {Simard}, L. and {White}, S. and {Zaritsky}, D.},
        title = "{Weak lensing mass reconstructions of the ESO Distant Cluster Survey}",
      journal = {\aap},
         year = 2006,
        month = may,
       volume = {451},
       number = {2},
        pages = {395-408},
          doi = {10.1051/0004-6361:20041787},
archivePrefix = {arXiv},
       eprint = {astro-ph/0511746},
 primaryClass = {astro-ph},
       adsurl = {https://ui.adsabs.harvard.edu/abs/2006A&A...451..395C}
}

@ARTICLE{2019A&A...628A.100D,
       author = {{Di Mascolo}, Luca and {Mroczkowski}, Tony and {Churazov}, Eugene and {Markevitch}, Maxim and {Basu}, Kaustuv and {Clarke}, Tracy E. and {Devlin}, Mark and {Mason}, Brian S. and {Randall}, Scott W. and {Reese}, Erik D. and {Sunyaev}, Rashid and {Wik}, Daniel R.},
        title = "{An ALMA+ACA measurement of the shock in the Bullet Cluster}",
      journal = {\aap},
         year = 2019,
        month = aug,
       volume = {628},
          eid = {A100},
        pages = {A100},
          doi = {10.1051/0004-6361/201936184},
archivePrefix = {arXiv},
       eprint = {1907.07680},
 primaryClass = {astro-ph.CO},
       adsurl = {https://ui.adsabs.harvard.edu/abs/2019A&A...628A.100D}
}

@ARTICLE{Ogrean2015,
       author = {{Ogrean}, G.~A. and {van Weeren}, R.~J. and {Jones}, C. and {Clarke}, T.~E. and {Sayers}, J. and {Mroczkowski}, T. and {Nulsen}, P.~E.~J. and {Forman}, W. and {Murray}, S.~S. and {Pandey-Pommier}, M. and {Randall}, S. and {Churazov}, E. and {Bonafede}, A. and {Kraft}, R. and {David}, L. and {Andrade-Santos}, F. and {Merten}, J. and {Zitrin}, A. and {Umetsu}, K. and {Goulding}, A. and {Roediger}, E. and {Bagchi}, J. and {Bulbul}, E. and {Donahue}, M. and {Ebeling}, H. and {Johnston-Hollitt}, M. and {Mason}, B. and {Rosati}, P. and {Vikhlinin}, A.},
        title = "{Frontier Fields Clusters: Chandra and JVLA View of the Pre-merging Cluster MACS J0416.1-2403}",
      journal = {\apj},
         year = 2015,
        month = oct,
       volume = {812},
       number = {2},
          eid = {153},
        pages = {153},
          doi = {10.1088/0004-637X/812/2/153},
archivePrefix = {arXiv},
       eprint = {1505.05560},
 primaryClass = {astro-ph.CO},
       adsurl = {https://ui.adsabs.harvard.edu/abs/2015ApJ...812..153O}
}

@ARTICLE{2022AstL...48..702B,
       author = {{Burenin}, R.~A. and {Zaznobin}, I.~A. and {Medvedev}, P.~S. and {Gilfanov}, M.~R. and {Kotov}, S.~S. and {Uklein}, R.~I. and {Dodonov}, S.~N. and {Moiseev}, A.~V. and {Eselevich}, M.~V. and {Bikmaev}, I.~F. and {Lyskova}, N.~S. and {Meshcheryakov}, A.~V. and {Postnov}, K.~A. and {Sazonov}, S. Yu. and {Starobinsky}, A.~A. and {Sunyaev}, R.~A. and {Churazov}, E.~M.},
        title = "{Observations of Massive Galaxy Clusters from the All-Sky Survey with the eROSITA Telescope Onboard the SRG Space Observatory}",
      journal = {Astronomy Letters},
         year = 2022,
        month = dec,
       volume = {48},
       number = {12},
        pages = {702-723},
          doi = {10.1134/S1063773723010012},
       adsurl = {https://ui.adsabs.harvard.edu/abs/2022AstL...48..702B}
}

@ARTICLE{2005RvMP...77..207V,
       author = {{Voit}, G. Mark},
        title = "{Tracing cosmic evolution with clusters of galaxies}",
      journal = {Reviews of Modern Physics},
         year = 2005,
        month = apr,
       volume = {77},
       number = {1},
        pages = {207-258},
          doi = {10.1103/RevModPhys.77.207},
archivePrefix = {arXiv},
       eprint = {astro-ph/0410173},
 primaryClass = {astro-ph},
       adsurl = {https://ui.adsabs.harvard.edu/abs/2005RvMP...77..207V}
}

@ARTICLE{2025A&A...694A.182A,
       author = {{Adam}, R. and {Eynard-Machet}, T. and {Bartalucci}, I. and {Cherouvrier}, D. and {Clerc}, N. and {Di Mascolo}, L. and {Dupourqu{\'e}}, S. and {Ferrari}, C. and {Mac{\'\i}as-P{\'e}rez}, J.-F. and {Pointecouteau}, E. and {Pratt}, G.~W.},
        title = "{PITSZI: Probing intra-cluster medium turbulence with Sunyaev─Zel'dovich imaging: Application to the triple merging cluster MACS J0717.5+3745}",
      journal = {\aap},
         year = 2025,
        month = feb,
       volume = {694},
          eid = {A182},
        pages = {A182},
          doi = {10.1051/0004-6361/202452342},
archivePrefix = {arXiv},
       eprint = {2409.14804},
 primaryClass = {astro-ph.CO},
       adsurl = {https://ui.adsabs.harvard.edu/abs/2025A&A...694A.182A}
}

@ARTICLE{2020NewAR..8801539H,
       author = {{Hardcastle}, M.~J. and {Croston}, J.~H.},
        title = "{Radio galaxies and feedback from AGN jets}",
      journal = {\nar},
         year = 2020,
        month = jun,
       volume = {88},
          eid = {101539},
        pages = {101539},
          doi = {10.1016/j.newar.2020.101539},
archivePrefix = {arXiv},
       eprint = {2003.06137},
 primaryClass = {astro-ph.HE},
       adsurl = {https://ui.adsabs.harvard.edu/abs/2020NewAR..8801539H}
}

@ARTICLE{2019A&A...622A..12H,
       author = {{Hardcastle}, M.~J. and {Williams}, W.~L. and {Best}, P.~N. and {Croston}, J.~H. and {Duncan}, K.~J. and {R{\"o}ttgering}, H.~J.~A. and {Sabater}, J. and {Shimwell}, T.~W. and {Tasse}, C. and {Callingham}, J.~R. and {Cochrane}, R.~K. and {de Gasperin}, F. and {G{\"u}rkan}, G. and {Jarvis}, M.~J. and {Mahatma}, V. and {Miley}, G.~K. and {Mingo}, B. and {Mooney}, S. and {Morabito}, L.~K. and {O'Sullivan}, S.~P. and {Prandoni}, I. and {Shulevski}, A. and {Smith}, D.~J.~B.},
        title = "{Radio-loud AGN in the first LoTSS data release. The lifetimes and environmental impact of jet-driven sources}",
      journal = {\aap},
         year = 2019,
        month = feb,
       volume = {622},
          eid = {A12},
        pages = {A12},
          doi = {10.1051/0004-6361/201833893},
archivePrefix = {arXiv},
       eprint = {1811.07943},
 primaryClass = {astro-ph.GA},
       adsurl = {https://ui.adsabs.harvard.edu/abs/2019A&A...622A..12H}
}

@ARTICLE{2026JHEAp..5300632Z,
       author = {{Zaznobin}, Igor and {Lyskova}, Natalya and {Bikmaev}, Ilfan and {Burenin}, Rodion and {Arshinova}, Arina and {Churazov}, Eugene and {Dodonov}, Sergey and {Gilfanov}, Marat and {Khabibullin}, Ildar and {Khamitov}, Irek and {Kotov}, Sergey and {Moiseev}, Alexey and {Sazonov}, Sergey and {Sunyaev}, Rashid and {Suslikov}, Mikhail and {Uklein}, Roman},
        title = "{Spectroscopic galaxy redshifts in the Peanut cluster ─ a massive nearly head-on cluster merger shortly after pericenter passage}",
      journal = {Journal of High Energy Astrophysics},
         year = 2026,
        month = jul,
       volume = {53},
          eid = {100632},
        pages = {100632},
          doi = {10.1016/j.jheap.2026.100632},
archivePrefix = {arXiv},
       eprint = {2603.10849},
 primaryClass = {astro-ph.GA},
       adsurl = {https://ui.adsabs.harvard.edu/abs/2026JHEAp..5300632Z}
}

@ARTICLE{2025RvMP...97d5004A,
       author = {{Adhikari}, Susmita and {Banerjee}, Arka and {Boddy}, Kimberly K. and {Cyr-Racine}, Francis-Yan and {Desmond}, Harry and {Dvorkin}, Cora and {Jain}, Bhuvnesh and {Kahlhoefer}, Felix and {Kaplinghat}, Manoj and {Nierenberg}, Anna and {Peter}, Annika H.~G. and {Robertson}, Andrew and {Sakstein}, Jeremy and {Zavala}, Jes{\'u}s},
        title = "{Astrophysical tests of dark matter self-interactions}",
      journal = {Reviews of Modern Physics},
         year = 2025,
        month = oct,
       volume = {97},
       number = {4},
          eid = {045004},
        pages = {045004},
          doi = {10.1103/m2vm-59y3},
archivePrefix = {arXiv},
       eprint = {2207.10638},
 primaryClass = {astro-ph.CO},
       adsurl = {https://ui.adsabs.harvard.edu/abs/2025RvMP...97d5004A}
}

@ARTICLE{2012ARA&A..50..353K,
       author = {{Kravtsov}, Andrey V. and {Borgani}, Stefano},
        title = "{Formation of Galaxy Clusters}",
      journal = {\araa},
         year = 2012,
        month = sep,
       volume = {50},
        pages = {353-409},
          doi = {10.1146/annurev-astro-081811-125502},
archivePrefix = {arXiv},
       eprint = {1205.5556},
 primaryClass = {astro-ph.CO},
       adsurl = {https://ui.adsabs.harvard.edu/abs/2012ARA&A..50..353K}
}

@preamble{ " \newcommand{\noop}[1]{} " }

@ARTICLE{1972CoASP...4..173S,
       author = {{Sunyaev}, R.~A. and {Zeldovich}, Ya. B.},
        title = "{The Observations of Relic Radiation as a Test of the Nature of X-Ray Radiation from the Clusters of Galaxies}",
      journal = {Comments on Astrophysics and Space Physics},
         year = 1972,
        month = nov,
       volume = {4},
        pages = {173},
       adsurl = {https://ui.adsabs.harvard.edu/abs/1972CoASP...4..173S}
}

@ARTICLE{2021A&A...656A.132S,
       author = {{Sunyaev}, R. and {Arefiev}, V. and {Babyshkin}, V. and {Bogomolov}, A. and {Borisov}, K. and {Buntov}, M. and {Brunner}, H. and {Burenin}, R. and {Churazov}, E. and {Coutinho}, D. and {Eder}, J. and {Eismont}, N. and {Freyberg}, M. and {Gilfanov}, M. and {Gureyev}, P. and {Hasinger}, G. and {Khabibullin}, I. and {Kolmykov}, V. and {Komovkin}, S. and {Krivonos}, R. and {Lapshov}, I. and {Levin}, V. and {Lomakin}, I. and {Lutovinov}, A. and {Medvedev}, P. and {Merloni}, A. and {Mernik}, T. and {Mikhailov}, E. and {Molodtsov}, V. and {Mzhelsky}, P. and {M{\"u}ller}, S. and {Nandra}, K. and {Nazarov}, V. and {Pavlinsky}, M. and {Poghodin}, A. and {Predehl}, P. and {Robrade}, J. and {Sazonov}, S. and {Scheuerle}, H. and {Shirshakov}, A. and {Tkachenko}, A. and {Voron}, V.},
        title = "{SRG X-ray orbital observatory. Its telescopes and first scientific results}",
      journal = {\aap},
         year = 2021,
        month = dec,
       volume = {656},
          eid = {A132},
        pages = {A132},
          doi = {10.1051/0004-6361/202141179},
archivePrefix = {arXiv},
       eprint = {2104.13267},
 primaryClass = {astro-ph.HE},
       adsurl = {https://ui.adsabs.harvard.edu/abs/2021A&A...656A.132S}
}

@ARTICLE{2021A&A...647A...1P,
       author = {{Predehl}, P. and {Andritschke}, R. and {Arefiev}, V. and {Babyshkin}, V. and {Batanov}, O. and {Becker}, W. and {B{\"o}hringer}, H. and {Bogomolov}, A. and {Boller}, T. and {Borm}, K. and {Bornemann}, W. and {Br{\"a}uninger}, H. and {Br{\"u}ggen}, M. and {Brunner}, H. and {Brusa}, M. and {Bulbul}, E. and {Buntov}, M. and {Burwitz}, V. and {Burkert}, W. and {Clerc}, N. and {Churazov}, E. and {Coutinho}, D. and {Dauser}, T. and {Dennerl}, K. and {Doroshenko}, V. and {Eder}, J. and {Emberger}, V. and {Eraerds}, T. and {Finoguenov}, A. and {Freyberg}, M. and {Friedrich}, P. and {Friedrich}, S. and {F{\"u}rmetz}, M. and {Georgakakis}, A. and {Gilfanov}, M. and {Granato}, S. and {Grossberger}, C. and {Gueguen}, A. and {Gureev}, P. and {Haberl}, F. and {H{\"a}lker}, O. and {Hartner}, G. and {Hasinger}, G. and {Huber}, H. and {Ji}, L. and {Kienlin}, A. v. and {Kink}, W. and {Korotkov}, F. and {Kreykenbohm}, I. and {Lamer}, G. and {Lomakin}, I. and {Lapshov}, I. and {Liu}, T. and {Maitra}, C. and {Meidinger}, N. and {Menz}, B. and {Merloni}, A. and {Mernik}, T. and {Mican}, B. and {Mohr}, J. and {M{\"u}ller}, S. and {Nandra}, K. and {Nazarov}, V. and {Pacaud}, F. and {Pavlinsky}, M. and {Perinati}, E. and {Pfeffermann}, E. and {Pietschner}, D. and {Ramos-Ceja}, M.~E. and {Rau}, A. and {Reiffers}, J. and {Reiprich}, T.~H. and {Robrade}, J. and {Salvato}, M. and {Sanders}, J. and {Santangelo}, A. and {Sasaki}, M. and {Scheuerle}, H. and {Schmid}, C. and {Schmitt}, J. and {Schwope}, A. and {Shirshakov}, A. and {Steinmetz}, M. and {Stewart}, I. and {Str{\"u}der}, L. and {Sunyaev}, R. and {Tenzer}, C. and {Tiedemann}, L. and {Tr{\"u}mper}, J. and {Voron}, V. and {Weber}, P. and {Wilms}, J. and {Yaroshenko}, V.},
        title = "{The eROSITA X-ray telescope on SRG}",
      journal = {\aap},
         year = 2021,
        month = mar,
       volume = {647},
          eid = {A1},
        pages = {A1},
          doi = {10.1051/0004-6361/202039313},
archivePrefix = {arXiv},
       eprint = {2010.03477},
 primaryClass = {astro-ph.HE},
       adsurl = {https://ui.adsabs.harvard.edu/abs/2021A&A...647A...1P}
}

@ARTICLE{1999PhR...310...97B,
       author = {{Birkinshaw}, M.},
        title = "{The Sunyaev-Zel'dovich effect}",
      journal = {\physrep},
         year = 1999,
        month = mar,
       volume = {310},
       number = {2-3},
        pages = {97-195},
          doi = {10.1016/S0370-1573(98)00080-5},
archivePrefix = {arXiv},
       eprint = {astro-ph/9808050},
 primaryClass = {astro-ph},
       adsurl = {https://ui.adsabs.harvard.edu/abs/1999PhR...310...97B}
}
\clearpage
\begin{appendix}
\section{LOFAR HBA image}
In Figure\,\ref{fig::HBA}, we show the LOFAR high-resolution image of the Peanut cluster. The image has a restoring beam size of $13\arcsec \times 4.6\arcsec$ and an rms of $200\,\mu \rm Jy\, beam^{-1}$. The halo emission is not detected in the available LOFAR HBA data owing to limitations in data quality.

\begin{figure}[!ht]
  \centering
  \includegraphics[width=\columnwidth]{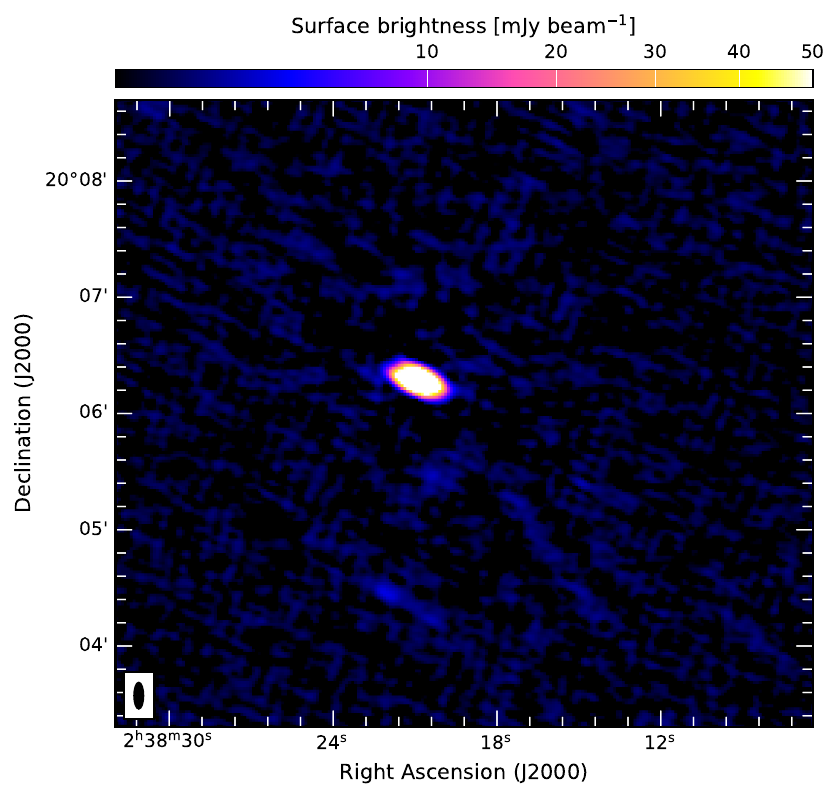}
  \caption{LOFAR HBA 144~MHz image of the Peanut cluster showing the bright double-lobe source. The image is a square-root intensity scale, and the beam size is indicated in the bottom left corner.}
  \label{fig::HBA}
\end{figure}

\section{Spectral index error map}
\label{sec::index_error}
In Figure\,\ref{fig::spec_error}, we show the spectral index error map corresponding to Figure\,\ref{fig::spec} (right panel), at a resolution of 15\arcsec. The uncertainty is $\approx0.2$ in the brightest regions of the halo and increases to $\approx0.6-0.7$ toward the faint outer regions.

\begin{figure}[!ht]
  \centering
  \includegraphics[width=\columnwidth]{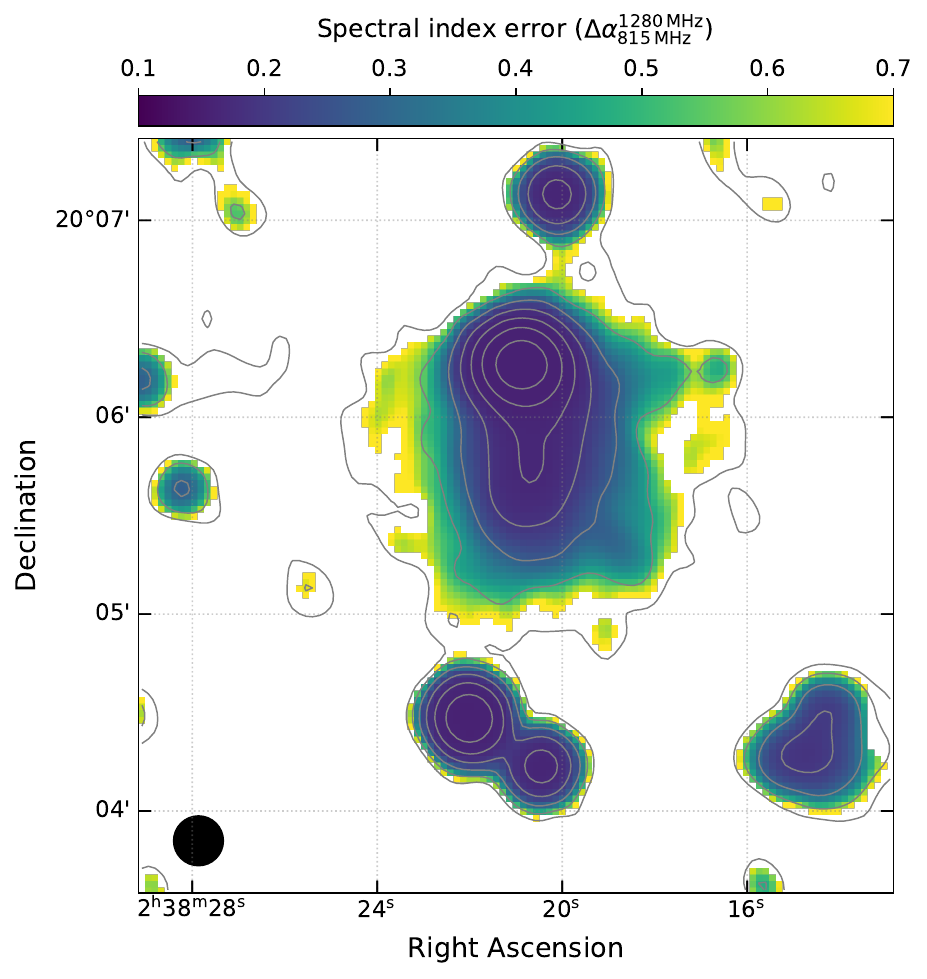}
  \caption{Spectral index error map between 815~MHz and 1.28~GHz at $15\arcsec$ resolution, corresponding to the spectral index map shown in Figure\,\ref{fig::spec} right panel. Overlaid are the MeerKAT L-band contours drawn at $[1, 2, 4, 8 ...]\times 3.0\sigma_{\rm rms}$. The beam size is indicated in the bottom left corner.}
  \label{fig::spec_error}
\end{figure}

\section{Cells used for the radio--X-ray comparison}
\label{sec::irx_cells}
Figure\,\ref{fig::IRX_cells} shows the grid of $15\arcsec$ cells used for the point-to-point radio--X-ray surface brightness analysis described in Sect.~\ref{sec::analysis}, overlaid on the MeerKAT UHF image. Cells with radio detections are drawn with solid lines, while cells with a radio surface brightness below $2\sigma$ are included as upper limits and shown with dashed lines. Cells covering the double-lobe source and other discrete sources were excluded.

\begin{figure}[!ht]
  \centering
  \includegraphics[width=\columnwidth]{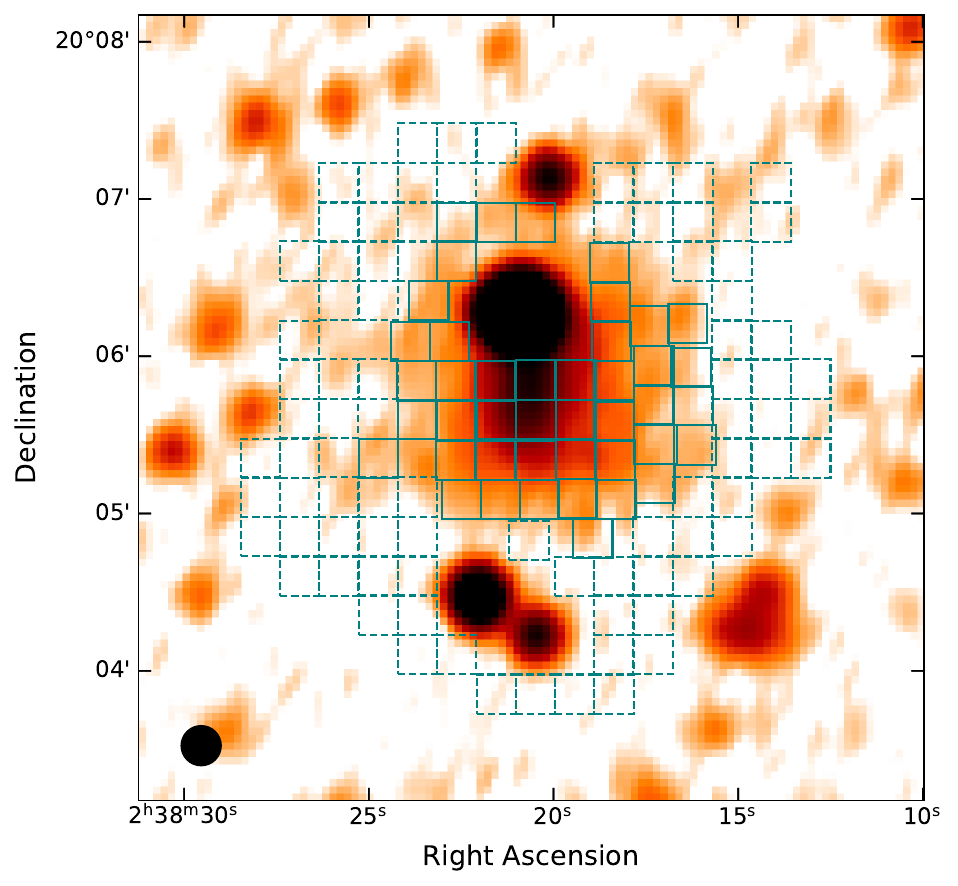}
  \caption{Grid of $15\arcsec$ cells used for the $I_{\rm R}-I_{\rm X}$ comparison, shown on the MeerKAT UHF image at $15\arcsec$ resolution. Cells with radio detections are outlined by solid lines, while dashed lines indicate cells included as $2\sigma$ upper limits. The beam size is indicated in the bottom left corner.}
  \label{fig::IRX_cells}
\end{figure}

\section{Source C}
\label{sec::sourceC}
Using the RACS and TGSS survey images, \citet{Lyskova2025} reported an arc-like diffuse source $\approx5\arcmin$ east of the cluster center. Figure\,\ref{fig::sourceC} shows this source in our uGMRT Band~4 image at $4.6\arcsec\times3.4\arcsec$ resolution, overlaid on a Pan-STARRS1 $r+i$ composite image. At this resolution, the radio emission is resolved into a compact core, coincident with an optical galaxy, and two narrow tails extending northward and westward from the core. This morphology is consistent with that of a tailed radio galaxy. The redshift of the host galaxy is currently unknown.

\begin{figure}[!ht]
  \centering
  \includegraphics[width=\columnwidth]{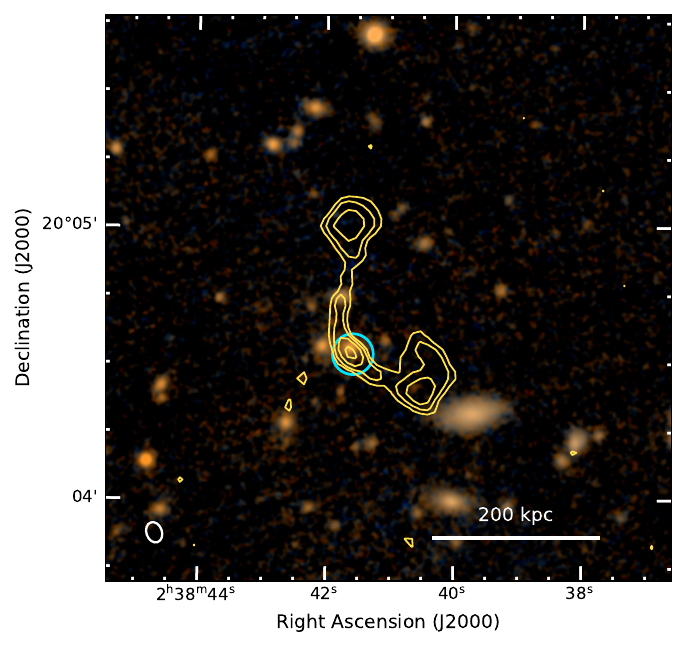}
  \caption{Source~C located $\approx5\arcmin$ east of the Peanut cluster center. The optical image is a Pan-STARRS1 $r+i$ composite and the contours are from the uGMRT Band~4 image at $4.6\arcsec\times3.4\arcsec$ resolution, drawn at $[1, 2, 4, 8 ...]\times3.0\sigma_{\rm rms}$ with $\sigma_{\rm rms}=40\,\mu{\rm Jy\,beam^{-1}}$. The circle marks the optical galaxy coincident with the compact radio component. The beam is shown in the bottom left corner.}
  \label{fig::sourceC}
\end{figure}

\end{appendix}

\end{document}